\documentclass[letterpaper,11pt]{article}
\usepackage{caption}
\pdfoutput=1 

\usepackage[title]{appendix}
\usepackage{extarrows}
\usepackage{amssymb}
\usepackage{amsmath}
\usepackage{setspace}

\usepackage{amstext}
\usepackage{etoolbox}
\let\tinymatrix\smallmatrix

\patchcmd{\tinymatrix}{\scriptstyle}{\scriptscriptstyle}{}{}
\patchcmd{\tinymatrix}{\scriptstyle}{\scriptscriptstyle}{}{}
\patchcmd{\tinymatrix}{\vcenter}{\vtop}{}{}
\patchcmd{\tinymatrix}{\bgroup}{\bgroup\scriptsize}{}{}

\usepackage{graphicx,epsfig}
\usepackage{epsfig}
\usepackage{verbatim} 
\usepackage{color}
\usepackage{ulem}
\usepackage{enumitem}
\usepackage{subfigure}
\usepackage{varwidth,xcolor}
\usepackage{bbm}
\usepackage{parskip}
\usepackage{dsfont}
\usepackage[numbers,sort&compress]{natbib}
\usepackage[body={6.5in, 9in}, right=1in, top=1in]{geometry}
\usepackage{mathtools}
\usepackage{comment}
\usepackage{caption}
\usepackage{bbold}
\usepackage{float}

\newcommand{\Comment}[1]{{}}
\definecolor{darkblue}{rgb}{0.15,0.35,0.55}
\definecolor{darkgreen}{rgb}{0.2,0.7,0.3}
\definecolor{reddish}{rgb}{0.65, 0.2, 0.2}
\usepackage[linktocpage=true]{hyperref}

\newcommand{\be}{\begin{equation}}
\newcommand{\ee}{\end{equation}}
\newcommand{\bea}{\begin{eqnarray}}
\newcommand{\eea}{\end{eqnarray}}
\newcommand{\beas}{\begin{eqnarray*}}
\newcommand{\eeas}{\end{eqnarray*}}

\def\({\left(}
\def\){\right)}

\newcommand\bigzero{\makebox(0,0){\text{\huge0}}}

\def\gsim{ \lower .75ex \hbox{$\sim$} \llap{\raise .27ex \hbox{$>$}} }
\def\lsim{ \lower .75ex \hbox{$\sim$} \llap{\raise .27ex \hbox{$<$}} }
\newcommand{\cev}[1]{\reflectbox{\ensuremath{\vec{\reflectbox{\ensuremath{#1}}}}}}

\usepackage{xcolor,colortbl}

\begin{document}
\def\thefootnote{\fnsymbol{footnote}}

\begin{center}
\LARGE{\textbf{Bayesian Reasoning in Eternal Inflation: A Solution to the Measure Problem}} \\[0.5cm]
 
\large{Justin Khoury and Sam S. C. Wong}
\\[0.5cm]

\small{
\textit{Center for Particle Cosmology, Department of Physics and Astronomy, University of Pennsylvania,\\ Philadelphia, PA 19104}}

\vspace{.2cm}

\end{center}

\vspace{.6cm}

\hrule \vspace{0.2cm}
\centerline{\small{\bf Abstract}}
{\small\noindent Probabilities in eternal inflation are traditionally defined as limiting frequency distributions, but a unique and
unambiguous probability measure remains elusive. In this paper, we present a different approach, based on Bayesian reasoning. 
Our starting point is the master equation governing vacuum dynamics, which describes a random walk on the network of vacua. Our probabilities require two pieces of prior information, both pertaining to initial conditions: a prior density~$\rho(t)$ for the time of nucleation, and a prior probability~$p_\alpha$ for the ancestral vacuum.
For ancestral vacua, we advocate the uniform prior as a conservative choice, though our conclusions are fairly insensitive to this choice.
For the time of nucleation, we argue that a uniform prior is consistent with the time-translational invariance of the master equation and represents the minimally-informative choice.
The resulting predictive probabilities coincide with Bousso's ``holographic" prior probabilities and are closely related to Garriga and Vilenkin's ``comoving" probabilities.
Despite making the least informative priors, these probabilities are surprisingly predictive. They favor vacua whose surrounding landscape topography is that of a deep funnel, akin to the folding funnels of naturally-occurring proteins. They predict that we exist during the approach to near-equilibrium, much earlier than the mixing time for the landscape. 
We also consider a volume-weighted~$\rho(t)$, which amounts to weighing vacua by physical volume. The predictive probabilities in this case coincide
with the GSVW measure. The Bayesian framework allows us to compare the plausibility of the uniform-time and volume-weighted hypotheses to explain our data
by computing the Bayesian evidence for each. We argue, under general and plausible assumptions, that posterior odds overwhelmingly favor the uniform-time hypothesis. 
\vspace{0.3cm}
\noindent
\hrule
\def\thefootnote{\arabic{footnote}}
\setcounter{footnote}{0}

\section{Introduction} 

Two separate developments in fundamental physics have led to the seemingly inescapable conclusion that our observable universe is part of a vast multiverse. The first development is the discovery 40 years ago of eternal inflation~\cite{Steinhardt:1982kg,Vilenkin:1983xq,Linde:1986fc,Linde:1986fd,Starobinsky:1986fx}. It is now understood that eternal inflation is a robust phenomenon that arises for a very wide class of scalar field potentials.
The second development is the realization that string theory admits a vast landscape of metastable states~\cite{Bousso:2000xa,Kachru:2003aw}. Much remains to be understood about the string landscape, in particular the subtle constraints that quantum gravity might place on low-energy physics~\cite{Obied:2018sgi,Agrawal:2018own,Palti:2019pca}, but it seems unambiguous that it is comprised of a myriad of 
metastable states, giving rise to a rich slew of effective field theories. Eternal inflation offers a mechanism to dynamically populate these vacua, resulting in the multiverse.

As an inhabitant of the multiverse, how should we reason probabilistically about the expected physical properties of our observable universe? Probabilities in eternal inflation are usually defined in terms of frequencies.
Concretely, the relative probabilities for two types of events~$A$ and~$B$ are defined as the ratio of their respective number of instances:
\be
\frac{P(A)}{P(B)} = \frac{N_A}{N_B} \,.
\label{prob freq}
\ee
The problem is that~$N_A$ and~$N_B$ are both infinite in an eternally-inflating universe, hence their ratio requires a regularization prescription (or ``measure") to be well-defined. This is usually done
by defining a finite space-time region with a geometric cutoff, evaluating~$N_A$ and~$N_B$ in this region, and taking the limit that the region becomes infinite. Unfortunately the result depends sensitively on the choice
of regulator. This is the measure problem. A variety of different proposals have been put forth over the years (see~\cite{Freivogel:2011eg} for a review). 

Aside from the issue of cutoff/regulator dependence, it seems that~\eqref{prob freq} is ambiguous even in a very large but finite universe. Consider the relative probabilities~$P(A)/P(B)$ to inhabit vacuum~$A$ or~$B$. It is unclear {\it a priori} whether one should count the number of bubbles of each type, the fraction of comoving or physical volume for each vacuum, or something else entirely. Simply put, what are the physical observables whose frequencies we are supposed to compare? 

In this paper we present an approach to the measure problem that instead relies on the Bayesian framework for probabilities. We adopt the notion of probability theory as an extension of deductive logic. As elegantly enunciated in the classic treatise by Jaynes~\cite{jaynes03}, this is the process of reasoning by which one draws uncertain conclusions (or ``reasonable expectations"~\cite{Cox}) from limited information. 

The Bayesian approach in eternal inflation is natural and fruitful for two reasons. First, as emphasized by Hartle and Srednicki~\cite{Hartle:2007zv,Srednicki:2009vb}, in a situation like eternal inflation where our data is replicated at (infinitely-many) other space-time locations, a prior assumption must be made about {\it our} location within the multiverse. Even if the fundamental theory handed us an unambiguous measure, such an assumption would still be necessary to translate third-person probabilities (what the measure predicts) to first-person probabilities (what {\it we} are most likely to observe). Thus one is inevitably led to work with first-person probabilities. A second, more practical reason is that all approaches to define a semi-classical measure (including ours) necessarily rely on certain working assumptions. The Bayesian scheme allows one to make all assumptions explicit through careful specification of prior information.

For simplicity and concreteness, we focus in this paper on false-vacuum eternal inflation, though our analysis carries over straightforwardly to stochastic (slow-roll) eternal inflation. Our starting point is the master equation governing the probabilities to occupy different vacua~\cite{Garriga:1997ef,Garriga:2005av}. This equation describes the Markov process along a given world-line for transitions between transient de Sitter (dS) vacua and into terminal Anti-de Sitter (AdS) or Minkowski vacua. The occupational probabilities are normalized at all times and are time-reparametrization invariant. 

A unique solution to the master equation requires specifying two pieces of prior information. The first piece of information is the {\it time of nucleation}. Since eternal inflation is geodesically past-incomplete~\cite{Borde:2001nh}, our bubble universe was nucleated a finite time~$t$ after the onset of inflation. The second piece of information is the initial condition. Along our world-line, false-vacuum eternal inflation started in some particular vacuum, which we call the {\it ancestral vacuum}~$\alpha$. Thus~$t$ and~$\alpha$ are model parameters for eternal inflation, and we must specify a prior probability distribution for each. {\it Different approaches to the measure problem amount to different choices for these priors.} Consistency requires that our priors reflect all information at hand, but should otherwise be minimally informative.

{\bf Prior for the ancestral vacuum:} The nature of the initial state in quantum cosmology has the subject of much debate over the years. Notable proposals include the Hartle-Hawking state~\cite{Hartle:1983ai}, the tunneling wave function~\cite{Vilenkin:1984wp,Vilenkin:1986cy}, and Linde's wavefunction~\cite{Linde:1983mx}. Arguments about the technical consistency of each proposal are still ongoing. At this point, even the basic question of whether the ancestral vacuum should preferentially have high-entropy/low-energy or low-entropy/high-energy remains unsettled. For our purposes, it therefore seems prudent to follow Laplace's {\it principle of indifference} and assign a uniform prior for ancestral vacua.

{\bf Prior for the time of nucleation:} Specifying a prior density~$\rho(t)$ is trickier, for the usual reason that an (improper) uniform prior for a continuous parameter is not reparametrization invariant.
However the symmetries of the problem at hand offer a natural solution. As argued by Jaynes~\cite{jaynes03}, consistency requires that~$\rho(t)$ should be invariant under all transformations
that leave our state of knowledge unchanged. Importantly, the master equation, from which our probabilities are derived, is invariant under time translations. A time translation can
justifiably have one of two effects on our state of knowledge:

\begin{enumerate}[label=(\roman*)]

\item It leaves our state of knowledge unchanged, reflecting complete ignorance about the time of nucleation. This leads us to impose an
(improper) uniform prior over proper (or e-folding) time, akin to a temporal Copernican principle.\footnote{A uniform prior over the real line is of course ill-defined, so we
will need to introduce a late-time cutoff. In the limit that the cutoff is removed, the posterior probabilities will be well-defined and time-reparametrization invariant.}  
The uniform-time distribution is the least informative prior. 

\item Alternatively, one might argue that a time translation increases the number of observers proportional to the growth in volume, hence~$\rho(t)$ should grow exponentially in time. This effectively
corresponds to weighing occupational probabilities by physical volume. For~$\rho(t)$ to be well-defined, it is necessary to impose a cutoff time~$t_{\rm c}$. The resulting prior density is
exponentially peaked near~$t_c$, to the extent that it can be well-approximated by a delta function.\footnote{This embodies a version of the ``youngness paradox"~\cite{Guth:2007ng}, though, as we will see, in our approach there is nothing paradoxical about this choice of prior.} Thus the late-time or volume-weighted prior is maximally informative.

\end{enumerate}  

These two prior densities reflect the assumptions implicit in most approaches to the measure problem. The late-time/volume-weighted prior is closely related to
measures based on the late-time, quasi-stationary distribution~\cite{Linde:1993nz,Linde:1993xx,GarciaBellido:1993wn,Vilenkin:1994ua,Garriga:1997ef,Garriga:2001ri,Garriga:2005av}. The assumption in this approach is that the evolution of the multiverse has been going on for an exponentially long time, much longer than the mixing time of the landscape. Correspondingly, we will find in Sec.~\ref{prior predict late-time sec} that the probabilities with late-time prior coincide with the quasi-stationary measure of Garriga, Vilenkin, Schwartz-Perlov and Winitzki (GSVW)~\cite{Garriga:2005av}. Because the prior peaks at late times, we refer to this case as the {\it late-time hypothesis}~${\cal H}_{\rm late}$.

The uniform-time prior, on the other hand, is representative of local measures, which focus on a space-time region around a time-like observer~\cite{Bousso:2006ev,Bousso:2009dm,Bousso:2010zi,Nomura:2011dt,Langhoff:2021uct}.
Because a typical world-line ends in a terminal vacuum well-before the mixing time for the landscape, the resulting probabilities differ markedly from the quasi-stationary distribution. Relatedly, an alternative framework developed recently~\cite{Khoury:2019yoo,Khoury:2019ajl,Kartvelishvili:2020thd,Khoury:2021grg,Khoury:2021zao} suggests that we live during the {\it approach to equilibrium}. (See~\cite{Denef:2017cxt} for related ideas.)
The resulting early-time measure~\cite{Khoury:2021grg} favors vacua that can be accessed early on in the evolution, instead of vacua that are frequently generated in the asymptotic future. We will show in Sec.~\ref{prior predictive with uniform time prior} that the probabilities with uniform-time prior coincide with Bousso's prior probabilities~\cite{Bousso:2006ev}\footnote{As explained in Sec.~\ref{prior predictive with uniform time prior}, Bousso first considers the ensemble of possible future ``histories" of a world-line. The probability of different histories coincides with our uniform-time predictive probabilities. Bousso then considers the causal diamond of each world-line in the ensemble, and calculates the fraction of observers making different observations within this causal diamond. Because we are interested in prior predictive probabilities, without any anthropic conditioning, this second aspect in Bousso's construction is not relevant for our purposes.}, and are closely related to Garriga and Vilenkin's ``comoving" probabilities~\cite{Garriga:2001ri,Garriga:2005av}. We refer to this case as the {\it uniform-time hypothesis}~${\cal H}_{\rm uni}$.

Our framework allow us to perform the three main operations of Bayesian inference.

\begin{enumerate}

\item By marginalizing over the model parameters~$t$ and~$\alpha$, we obtain in Sec.~\ref{prior predictive sec} the {\it prior predictive distributions}~$P(I|{\cal H}_i)$ for each hypothesis. These inform us on which vacua are statistically favored, without taking our data into consideration. They also allow us to compute probability distributions for various physical parameters, in particular the cosmological constant (CC).    

\item The uniform-time and late-time hypotheses will compete in Sec.~\ref{model comp} by computing the posterior odds ratio~$\frac{P({\cal H}_{\rm late}|D)}{P({\cal H}_{\rm uni}|D)}$. Assuming comparable prior odds for the two hypotheses, we will find, under general and reasonable assumptions about transition rates, that {\it posterior odds exponentially favor the uniform-time hypothesis}. That is, the GSVW measure~\cite{Garriga:2005av} is exponentially disfavored compared to the holographic prior probabilities~\cite{Bousso:2006ev}.  This is a key result of our analysis. There are possible caveats and loopholes in our analysis, and we will carefully spell these out in Sec.~\ref{loopholes}.

\item Conditioning on our data~$D$, we will perform parameter inference in Sec.~\ref{time of existence sec}. We will be specifically interested in the posterior probability distribution~$P(t|D,{\cal H}_i)$ for the time of nucleation. Focusing on~${\cal H}_{\rm uni}$, we will find that the average time for occupying vacua compatible with our data is much shorter than mixing time, confirming the assumptions underlying the early-time approach~\cite{Khoury:2019yoo,Khoury:2019ajl,Kartvelishvili:2020thd,Khoury:2021grg,Khoury:2021zao}. 

\end{enumerate}

We believe that the uniform-time measure is the correct objective approach to inductive reasoning in the multiverse. It consistently reflects our current state of knowledge about the ancestral vacuum and time of
nucleation --- we simply do not know at this point how/when eternal inflation started in our past. Importantly, despite adopting the least informative priors, the resulting measure is surprisingly predictive. It favors vacua that can be accessed through a sequence of downward transitions, from a large basin of high-energy vacua. Thus such vacua belong to a landscape region with the topography of a funnel~\cite{Khoury:2019yoo,Khoury:2019ajl,Khoury:2021grg}, akin to folding funnels of proteins~\cite{proteins1}. Furthermore, by predicting that we exist at early times in eternal inflation, the measure implies that we are ``normal" observers as opposed to freak observers (Boltzmann brains)~\cite{Albrecht:2002uz,Dyson:2002pf,Albrecht:2004ke,Page:2005ur,Page:2006dt} produced
on exponentially longer time scales. 

Our approach to the measure problem does not require ad hoc geometric constructions, nor are we counting anything. In phrasing the problem in terms of a prior~$\rho(t)$ for the time of existence,
we were inspired by Caves' elegant approach~\cite{Caves:2000tx} to the so-called Doomsday paradox~\cite{doom1,doom2,doom3} using similar Bayesian reasoning. Ultimately, despite the conceptual minefield inherent to eternal inflation, landscape dynamics reduce after suitable coarse-graining to a linear Markov process, {\it i.e.}, a random walk on the network of vacua. Such a mathematically simple problem ought to have a simple answer. Indeed, the probabilities obtained with the uniform-time prior are intuitively clear. They favor vacua that are easily accessed under the random walk.

In a forthcoming paper~\cite{future} we will show that the measure favors regions of the landscape that are close to the directed percolation phase transition. In other words, the measure selects regions of the landscape that are nearly tuned at criticality.\footnote{Interestingly, this is complementary to the mechanism of `self-organized localization'~\cite{Giudice:2021viw}, whereby the near-criticality of our universe arises from quantum first-order phase transitions in stochastic inflation. In contrast, our approach pertains to classical, second-order non-equilibrium criticality.} Furthermore, it translates to a probability distribution for the CC that favors a naturally small and positive vacuum energy.

Coincidentally with our paper, Ref.~\cite{Friedrich:2022tqk} appeared on the arXiv. Although their approach to the measure problem, based on the local Wheeler-De Witt equation, is quite different than the Bayesian method pursued here, the resulting measure appears to be quite similar. 

We close this Introduction with some brief remarks about anthropic reasoning. From a Bayesian perspective, anthropic reasoning is intermediate between prior predictive probabilities~$P(I|{\cal H})$, which are unconditioned, and posterior probabilities~$P({\cal H} | D)$, which are conditioned on our data. Anthropic conditionalization is an in-between~\cite{Aguirre:2004qb,Hartle:2004qv}, whereby one attempts to condition on the existence of observers. An immediate difficulty, of course, lies in defining ``observers" in sufficient generality,\footnote{We would be hard-pressed to offer such a definition within our observable universe, let alone across a multiverse with varying physical constants.} so one is forced in practice to condition on some observational proxy for the existence of observers. For instance, one approach is to consider $P(I|{\cal H}\cap{\cal A})$~\cite{Hartle:2004qv}, which are conditioned on the set of observational proxy~${\cal A}$ within the anthropic window. In this sense, within a hypothesis, only the set of universes within this anthropic window is considered. The problem is that the choice of proxy (fraction of baryons per galaxy~\cite{Weinberg:1987dv}, entropy production~\cite{Bousso:2006ev}, or number of observations~\cite{Starkman:2006at}) is ambiguous and can lead to very different results~\cite{Starkman:2006at}. For this reason, we avoid any anthropic conditionalization in this work. Our focus is on extracting as much information from prior predictive probabilities. 

\section{Vacuum dynamics as absorbing Markov process}
\label{RW complex nets}

The landscape can be modeled as a network (or graph) of nodes representing the various dS, AdS and Minkowski vacua. We assume as usual that AdS and Minkowski vacua are terminal, acting as absorbing nodes. Links define the network topology and represent all relevant transitions between vacua.\footnote{By ``relevant", we mean transitions with non-negligible rates on the time scale of interest.} In what follows, indices~$i,j,\ldots$ and~$a,b,\ldots$ denote dS and terminal vacua respectively, while capital indices~$I,J\ldots$ refer collectively to all vacua. Greek letters~$\alpha,\beta,\ldots$ denote the ancestral vacuum along our past world-line.

Following the seminal papers of Garriga, Vilenkin and collaborators~\cite{Garriga:1997ef,Garriga:2005av}, cosmological evolution on the landscape is described by a
Markov process. Because of terminals, this is technically an absorbing Markov process --- detailed balance is explicitly violated, hence the dynamics are out-of-equilibrium. 
Along a given world-line, the probability~$f_I(\tau_I)$ to occupy vacuum~$I$ as a function of the local proper time~$\tau_I$
satisfies the master equation 
\be
\Delta f_I = \sum_J  \left( \kappa^{\text{proper}}_{IJ} - \delta_{IJ} \sum_K \kappa^{\text{proper}}_{KJ} \right) \Delta\tau_J\, f_J \,,
\label{master0}
\ee
where $\kappa_{IJ}^{\text{proper}}$ is the $J \rightarrow I$ proper transition rate. This equation holds for any tunneling mechanism (Coleman-De Luccia (CDL)~\cite{Coleman:1977py,Callan:1977pt,Coleman:1980aw}, Hawking-Moss~\cite{Hawking:1981fz}, Brown-Teittleboim~\cite{Brown:1987dd}), but whether it applies to ``upward" transitions is an open question~\cite{Garriga:1997ef}. For most of our analysis we will remain agnostic about the nature of transition rates.

The master equation relies on coarse-graining over a time interval~$\Delta \tau_I$, which should be longer than any transient evolution between epochs of vacuum energy domination. Within our own bubble universe, for instance, this amounts to coarse-graining over at least the last 14 billion years of radiation and matter domination, until vacuum energy comes to completely dominate. On the other hand,~$\Delta \tau_I$ cannot be arbitrarily long.  
It should be shorter than the lifetime of most metastable dS vacua, for otherwise we would be ``integrating out" the transitions we are interested in describing. In practice, the coarse-graining time interval for a given transition to~$I$ should satisfy~$\Delta\tau_I \gtrsim |H_I|^{-1} \log \frac{H_{\rm parent}}{|H_I|}$, where~$H_{\rm parent}$ is the Hubble rate of the parent dS vacuum (see, {\it e.g.},~\cite{Salem:2012wa}). Thus it suffices to assume

\be
\Delta\tau_I = |H_I|^{-1} \log \frac{M_{\rm Pl}}{|H_I|}\,.
\label{coarsegrain}
\ee
In particular, since bubbles of AdS vacua crunch in a Hubble time, coarse-graining spans their entire evolution. An AdS bubble nucleated at a given time crunches and
dies within a time~$\Delta \tau_I$ later. 

It is convenient to define a general time variable~$t$, related to proper time via a lapse function:
\be
\Delta \tau_I = {\cal N}_I \Delta t\,. 
\ee
In terms of~$t$,~\eqref{master0} becomes
\be
\Delta f_I = \sum_J  \left( \kappa_{IJ} - \delta_{IJ} \sum_K \kappa_{KJ} \right) \Delta t\, f_J \,,
\label{master1}
\ee
where
\be
\kappa_{IJ} \equiv \kappa^{\text{proper}}_{IJ} {\cal N}_J\,. 
\label{kappa gen}
\ee
Equation~\eqref{master1} makes two properties of the~$f_I$'s manifestly clear:~$1)$~The master equation~\eqref{master1}
is manifestly invariant under redefinitions of~$t$, hence the~$f_I$'s are time-reparameterization invariant; $2)$~Because summing the right-hand side over~$I$ gives zero, the~$f_I$'s
can be normalized:~$\sum_I f_I = 1$. Thus the~$f_I(t)$'s give well-defined, gauge-invariant probabilities to occupy different vacua at time~$t$.

In the continuum limit ($\kappa_{IJ} \Delta t \ll 1$), this reduces to
\be
\frac{{\rm d}f_I}{{\rm d}t} = \sum_{J} \mathbb{M}_{IJ}f_{J} \,,
\label{master}
\ee
where~$\mathbb{M}_{IJ} \equiv \kappa_{IJ} - \delta_{IJ} \sum_K \kappa_{KJ}$ is the transition matrix. 
We will be primarily interested in the dS component of this equation, given by
\be
\frac{{\rm d}f_i}{{\rm d}t}  = \sum_j M_{ij} f_j  \,,
\label{master dS}
\ee
where 
\be
M_{ij}\equiv \kappa_{ij} - \delta_{ij} \kappa_j
\label{Mdef}
\ee
is the~${\rm dS}\rightarrow {\rm dS}$ transition matrix, and~$\kappa_i \equiv \sum_J \kappa_{Ji}$ is the total decay rate of vacuum~$i$. (Note that~$\kappa_i$ includes decay channels into dS as well as terminal vacua.) Our only assumption about~$M_{ij}$ is that it is irreducible, {\it i.e.}, there exists a sequence of transitions connecting any pair of dS vacua. This property has been argued to be valid for the string landscape~\cite{Brown:2011ry}.

Importantly, the form of~\eqref{master} and~\eqref{master dS} makes clear that the master equation is {\it time-translation invariant}. To be precise, it is invariant under translations of any time variable~$t$ related to proper time via a lapse function~${\cal N}_I$ that depends on~$H_I$ only. This includes proper time as well as e-folding time. Later on we will invoke this symmetry to justify the uniform-time prior.

\subsection{Green's function}

Equation~\eqref{master dS} can be solved in terms of a Green's function:
\be
f_i(t) = \sum_\alpha \left({\rm e}^{Mt}\right)_{i\alpha} p_\alpha\,, 
\label{f soln}
\ee
where~$p_\alpha \equiv f_\alpha(0)$ is the initial probability over ancestral vacua. 
Later on, we will need the Laplace transform of the Green's function:
\be
\int_0^\infty {\rm d}t \left({\rm e}^{Mt}\right)_{ij} {\rm e}^{-st} = (s - M)^{-1}_{ij} \,.
\label{Laplace}
\ee
Using~\eqref{Mdef}, this factorizes as 
\be
(s - M)^{-1}_{ij} = (s+\kappa_i)^{-1} \left(\mathds{1} - T(s)\right)^{-1}_{ij} \,;\qquad T_{ij}(s) \equiv \frac{\kappa_{ij}}{s+ \kappa_j} \,.
\label{Green}
\ee
In particular,~$T_{ij} \equiv T_{ij}(0) =  \frac{\kappa_{ij}}{\kappa_j}$ is the branching ratio. 

The matrix~$\left(\mathds{1} - T\right)^{-1}$ is known as the {\it fundamental matrix} for the absorbing Markov chain.
In the theory of Markov chains,~$\left(\mathds{1} - T\right)^{-1}_{ij}$ gives the expected number of visits to~$i$ starting from~$j$ before reaching terminals.\footnote{In particular,~$\sum\limits_i\left(\mathds{1} - T\right)^{-1}_{ij}$ gives the expected number of steps before reaching terminals starting from $j$.}
This is easily seen by expanding it as a geometric series,~$\left(\mathds{1} - T\right)^{-1} =\mathds{1} + T + T^2 + \ldots$,
and recognizing that the~$n^{\rm th}$ term in the series,~$\left(T^n\right)_{ij}$, represents a branching probability for the $n$-step chain~$j \rightarrow \ldots \rightarrow i$,
summed over all intermediaries. In other words,
\be
\left(\mathds{1} - T\right)^{-1}_{ij} = \sum\limits_{\substack{\text{paths} \\ j\rightarrow i}} \eta_{ij}\,, 
\label{T branch}
\ee
where~$\eta \equiv \prod_{\rm edges} T$ is the branching probability for each path connecting~$j$ to~$i$. For completeness, let us also define the full branching ratio matrix $\mathbb{T}_{IJ}$ as follows:
\be
\mathbb{T}_{ij} =T_{ij} = \frac{\kappa_{ij}}{\kappa_j}\,;\qquad \mathbb{T}_{Ij} =  \frac{\kappa_{Ij}}{\kappa_j}\,;\qquad \mathbb{T}_{ab} = \delta_{ab}\,;\qquad \mathbb{T}_{ja} = 0\,.
\label{branch mtx}
\ee
This ensures that~$\mathbb{T}_{IJ}$ satisfies, for all~$J$,
\be  
 \sum_{I} \mathbb{T}_{IJ}=1 \,.
\label{branch mtx sum}
\ee

\subsection{Detailed balance and downward approximation}
\label{down sec}

Most of our analysis will hold for general transition rates between vacua. To simplify some of the expressions below, however, it will be convenient at some point to make a
very general and reasonable assumption about these rates, namely that transitions between dS vacua satisfy a condition of {\it detailed balance}~\cite{Lee:1987qc}:
\be
\frac{\kappa_{ji}}{\kappa_{ij}}  \sim {\rm e}^{S_j-S_i}\,,
\label{detailed balance}
\ee
where~$S_j =  \frac{8\pi^2M_{\rm Pl}^2}{H_j^2}$ is the dS entropy. This assumption is satisfied by CDL, Hawking-Moss and Brown-Teittleboim tunneling, and has been oft-invoked
in earlier works on the measure problem. Notably it is violated by the Farvi-Guth-Guven process~\cite{Farhi:1989yr}, though the interpretation of its
singular instanton remains an open question~\cite{Fischler:1989se,Fischler:1990pk,DeAlwis:2019rxg,Fu:2019oyc,Mirbabayi:2020grb}. It is also violated by the mechanism of nucleating localized,
high-energy regions proposed recently~\cite{Olum:2021pux}.

Equation~\eqref{detailed balance} implies that upward transitions, which increase the potential energy, are exponentially suppressed compared to downward tunneling.
This allows one to define a ``downward" approximation~\cite{SchwartzPerlov:2006hi,Olum:2007yk}, in which upward transitions are treated perturbatively. 
Labeling dS vacua for convenience in order of increasing potential energy,~$0 < V_1 \leq \ldots \leq V_{N_{\rm dS}}$, the transition matrix becomes
upper-triangular to zeroth order in this approximation:
\be
M \simeq \begin{bmatrix}
-\kappa_1 & \kappa_{12} & \kappa_{13} &  \ldots  \\
 & -\kappa_2 & \kappa_{23} &  \ldots \\
  & & -\kappa_3  & \ldots \\
 &  \raisebox{3ex}[0ex][0ex]{\bigzero} &  & \ddots
\end{bmatrix}  \,.
\label{M original}
\ee
Thus the eigenvalues at this order are simply by given by its diagonal entries, {\it i.e.}, by the decay rates~$\kappa_i$ of individual vacua. In particular, the largest (least negative) eigenvalue~$\lambda_1$ is set by the most stable vacuum, also known as the {\it dominant vacuum}~$\star$:\footnote{It is conceivable that the landscape features a cluster of dominant vacua, with nearly degenerate decay rates. We ignore this possibility for simplicity.}
\be
q \equiv -\lambda_1 \simeq \kappa_\star \,.
\label{q def}
\ee
More generally, it can be shown rigorously that~$q\leq \kappa_\star$~\cite{Garriga:2005av}. The corresponding dominant eigenvector, denoted by~$s_j$, also admits a simple perturbative expression~\cite{Olum:2007yk}:
\be
s_j = \delta_{j\star} + \frac{\kappa_{j\star}}{\kappa_j - \kappa_\star} + \sum_{\ell \neq \star} \frac{\kappa_{j\ell}\kappa_{\ell\star}}{(\kappa_j - \kappa_\star)(\kappa_\ell - \kappa_\star)}+\ldots
\label{s pert}
\ee
Assuming that~$\kappa_j \gg \kappa_\star$ for all~$j\neq \star$, which is reasonable since rates are typically exponentially staggered, the series can be resummed compactly as
\be
s_j \simeq \frac{\kappa_\star}{\kappa_j} \left(\mathds{1} - T\right)^{-1}_{j\star} \,.
\label{s simple}
\ee
Following~\cite{Olum:2007yk}, we have neglected sequences of transitions which return to~$\star$ at least once before reaching~$j$, since these involve additional upward transitions
and therefore amount to exponentially small corrections.

\subsection{First-passage statistics}

First-passage statistics~\cite{Redner} offer a useful tool to study false-vacuum eternal inflation~\cite{Khoury:2019yoo,Khoury:2019ajl,Kartvelishvili:2020thd,Khoury:2021grg}, and have
also been applied to stochastic inflation~\cite{Vennin:2015hra,Assadullahi:2016gkk,Vennin:2016wnk,Noorbala:2018zlv}. We briefly review a few elementary results that will be
helpful in our analysis. 

The first-passage density,~$F_{ij}(t)$~($i\neq j$), is defined as the probability density that a random walker starting from~$j$ visits~$i$ for the first
time at time~$t$.\footnote{In this paper we will only need~${\rm dS}\rightarrow {\rm dS}$ first-passage statistics. See~\cite{Khoury:2019ajl} for~${\rm dS}\rightarrow {\rm AdS}$ results.} 
The first-passage density is related to the Green's function through the well-known result~\cite{Redner}:
\be
\left({\rm e}^{Mt}\right)_{ij} = \int_{0^-}^t {\rm d}t'\, \left({\rm e}^{M(t-t')}\right)_{ii}  F_{ij}(t') \,.
\label{renewal}
\ee
This equation is valid for all~$t \geq 0$ for $i\neq j$, and for all~$t>0$ for $i = j$ with $F_{ii}(t')=\delta(t')$. Thus the occupational probability at time~$t$ is the
probability of reaching~$i$ for the first time at any earlier time~$t'$, multiplied by the loop probability for returning to~$i$ in the remaining time. For $i \neq j$, the solution is
given in terms of Laplace transforms:
\be
\tilde{F}_{ij}(s) = \frac{(s-M)^{-1}_{ij}}{(s-M)^{-1}_{ii}} = \frac{\left(\mathds{1} - T(s)\right)^{-1}_{ij}}{\left(\mathds{1} - T(s)\right)^{-1}_{ii}}\,,
\label{F(s)}
\ee
where we have used~\eqref{Green}. 

All first-passage statistics can be derived from~$F$. For instance, its integral gives the ever-hitting probability:
\be
{\cal P}_{ij}  \equiv \int_0^\infty {\rm d}t \, F_{ij}(t) = \tilde{F}_{ij}(0) = \frac{\left(\mathds{1} - T\right)^{-1}_{ij}}{\left(\mathds{1} - T \right)^{-1}_{ii}} \,; \qquad (i\neq j)\,.
\label{first passage prob}
\ee
Note that this probability is less than unity because of terminals. Meanwhile, its first moment gives the mean first-passage time (MFPT):
\be
\langle t_{ij}\rangle \equiv \frac{1}{{\cal P}_{ij}} \int_0^\infty {\rm d}t \,t F_{ij}(t) = - \left.\frac{{\rm d}\ln \tilde{F}_{ij}(s)}{{\rm d}s}\right\vert_{s=0}  \,; \qquad (i\neq j)\,.
\label{MFPT def}
\ee
This gives the average time starting from~$j$ and conditioned on hitting the target~$i$. Similar expressions can be obtained for the ever-return probability~${\cal P}_{ii}$ and mean first-return time~$\langle t_{ii}\rangle$~\cite{Khoury:2019ajl}. These expressions simplify in the downward approximation. To zeroth order in this approximation, the network becomes
acyclic, and therefore~$\left(\mathds{1} - T(s)\right)^{-1}_{ii} \simeq 1$. It follows that
\be
{\cal P}_{ij} \simeq \left(\mathds{1} - T\right)^{-1}_{ij} \qquad (i\neq j)\,.
\label{P down}
\ee

\section{Bayesian reasoning in eternal inflation}

As a first step in defining probabilities, we must carefully distinguish the elements that are inherent to the eternal inflation hypothesis from those 
that require additional assumptions in the form of prior information. 

\begin{itemize}

\item Since eternal inflation is not past geodesically complete~\cite{Borde:2001nh}, it started a finite time~$t$ in our past. We do not know how long ago that was. 

\item Along our past world-line, false-vacuum eternal inflation, governed by the master equation~\eqref{master}, started out in some ancestral dS vacuum~$\alpha$. We do not know which one. 

\item Our bubble universe was nucleated at time~$t$ in some parent dS vacuum~$j$, but we do not know which one. We will condition our probabilities on one piece of data, namely that we exist in the transient period before vacuum domination, that is, within a coarse-graining time~$\Delta t$ after nucleation.

\end{itemize}

Our starting point is to define the conditional probability~$\wp(j | t,\alpha)$ to occupy parent vacuum~$j$ at given time~$t$, given an ancestral vacuum~$\alpha$.
The Green's function offers an unambiguous and gauge-invariant definition of this probability:
\be
\wp(j | t,\alpha) = \left({\rm e}^{Mt}\right)_{j\alpha} \,.
\label{parent cond}
\ee
The joint probability~$\wp(j,t,\alpha)$ is obtained as usual by multiplying with a prior probability distribution:
\be
\wp(j,t,\alpha) = \left({\rm e}^{Mt}\right)_{j\alpha} P_{\rm prior}(t,\alpha)\,.
\label{parent joint}
\ee
Since the time of nucleation and ancestral vacuum correspond {\it a priori} to logically distinct assumptions, it is natural to assume they are independent:
\be
P_{\rm prior}(t,\alpha) = p_\alpha \rho(t)\,.
\label{prior factorize}
\ee
The~$p_\alpha$'s, defined in~\eqref{f soln}, are discrete probabilities for ancestral vacua. They satisfy~$\sum\limits_{\alpha = 1}^{N_{\rm dS}} p_\alpha = 1$.
Meanwhile,~$\rho(t)$ is a prior probability density for the time of nucleation. It satisfies 
\be
\int_0^\infty {\rm d}t \,\rho(t)  = 1\,.
\ee
We will discuss at length two justified choices for~$\rho(t)$ in Sec.~\ref{time priors}. An analogous prior for the time of observation was discussed by Caves~\cite{Caves:2000tx} in a different context.

The distribution~\eqref{parent joint} pertains to parent dS vacua. The joint probability distribution~$P(I,t,\alpha)$ to inhabit vacuum~$I$ within~$\Delta t$ after a nucleation event at time~$t$, starting from~$\alpha$, is given by 
\bea
\nonumber
P(I,t,\alpha) &=& {\cal N} \sum_{j} \kappa_{Ij}\Delta t \, \wp(j,t,\alpha)   \\ 
&= & {\cal N} \sum_{j} \kappa_{Ij}\Delta t \,  \left({\rm e}^{Mt}\right)_{j\alpha} p_\alpha \rho(t)  \,.
\label{P cond}
\eea
As mentioned above, this probability distribution is conditioned on our bubble being nucleated within the last~$\Delta t$. The normalization constant~${\cal N}$ will be fixed shortly. It is important to stress that~$P(I,t,\alpha)$ and~$\wp(j,t,\alpha)$ are different probabilities, because the former assumes that~$I$ is reached within the last~$\Delta t$.\footnote{Relatedly, one may be tempted to infer from~\eqref{P cond} that~$P(I,t,\alpha) = P(I,\alpha | t)\rho(t)$, but this is incorrect. (For instance, the $P(I,\alpha | t)$'s thus defined would not be normalized over~$I$ and~$\alpha$.) Instead, the correct conditional probabilities are~$P(I,\alpha | t) = P(I,t,\alpha)/\sum\limits_{J,\beta} P(J,t,\beta)$.} Notice that~$I$ can be either dS or terminal, since we are not conditioning on our data at this stage.

\subsection{Prior predictive distribution}
\label{P(I)} 

Marginalizing over the model parameters~$t$ and $\alpha$ gives the {\it prior predictive distribution}:
\bea
\nonumber
P(I) &=& \sum_\alpha \int_0^\infty {\rm d} t  P(I,t,\alpha)  \\
&=& {\cal N} \sum_{j} \kappa_{Ij}\Delta t \int_0^\infty {\rm d} t \, f_j(t) \rho(t) \,,
\label{PI}
\eea
where we have used~\eqref{f soln}. Thus~$P(I)$ is the probability to inhabit vacuum~$I$, averaged over all possible time of existence and ancestral vacua. These probabilities inform us on which vacua are statistically favored, without taking our data into consideration (other than conditioning on our bubble being nucleated within the last~$\Delta t$). The normalization constant is fixed by demanding~$\sum_I P(I) = 1$. This yields the normalized probabilities
\be
\boxed{P(I)  = \frac{ \sum\limits_{j} \kappa_{Ij} \int\limits_0^\infty {\rm d} t\, f_j(t) \rho(t)}{\sum\limits_{k} \kappa_{k}   \int\limits_0^\infty {\rm d} t \, f_k(t) \rho(t)}} \,, 
\label{predictive distribution}
\ee
where we have used~$\sum_J \kappa_{Jk} = \kappa_k$. From the~$P(I)$'s, one can make predictions for various observables. For instance, the predictive probability density~$\rho(\Lambda)$ for the CC is obtained by summing over all vacua with vacuum energy between~$\Lambda$ and~$\Lambda + {\rm d}\Lambda$:
\be
\rho(\Lambda) {\rm d}\Lambda =  \sum_{I \,\in\, \Lambda < \Lambda_I < \Lambda + {\rm d}\Lambda}\,P(I)\,.
\ee

\subsection{Parameter inference}
\label{param infer}

A second operation of interest is to use our data~$D$ to infer the model parameters~$\alpha$ and~$t$. The data refers to all the information available about our observable universe, in the form
of measured values for various observables~$\{O_i\}$. These include the particle content, masses and couplings of the Standard Model, as well as the parameters of the cosmological~$\Lambda$CDM model. 
A vacuum compatible with our data should, in particular, have a suitable dark matter candidate with correct relic abundance, its vacuum energy should match the observed CC,\footnote{Hence we will only include dS vacua when summing over vacua compatible with~$D$.} and its surroundings should allow for a period of slow-roll inflation compatible with the observed amplitude and spectral tilt of primordial perturbations, {\it etc.} 

At a more philosophical level, there should be more layers of conditioning~\cite{Aguirre:2004qb}. For instance, the co-called observed Higgs mass of 125~GeV is not really a type of data~$D$ as the true form of data is simply the 125~GeV bump measured from a huge number of scattering events. The Standard Model including the Higgs boson, even quantum field theory, on the other hand, are in fact part of the hypothesis to be tested by these scattering events. Nevertheless, a confidence level of six sigma means that the hypothesis is verified to an extreme extent under this single measurement. Therefore we can just treat the 125~GeV boson as part of the data~$D$. Despite this, the remarkable accuracy still cannot not stop us from challenging the Standard Model by comparing it with alternatives through other measurements. When it comes to cosmology, hypothesis testing is far less accurate than that in particle scattering as we are limited to a single sample size, yet it is still possible to compare hypotheses using the same parameter inference strategy. Also, due to higher level of uncertainty, there is no reason not to think of alternatives to cosmological scenarios.

Parameter inference is performed using the posterior probability~$P(t,\alpha | D)$. We will be primarily interested in the dependence on the nucleation time, hence we will marginalize over ancestral vacua.
The posterior probability for the time of nucleation conditioned on our data is
\be
P(t | D ) =  \frac{ \sum\limits_{i \subset I_D} P(i,t)}{P(D)}  \,,
\label{Bayes}
\ee
where the denominator~$P(D) = \sum\limits_{j \subset I_D} P(j)$ is the probability of our data, and~$P(i, t) \equiv \sum\limits_\alpha  P(i, t, \alpha)$. To be clear, here~$I_D$ denotes the set of all vacua compatible with our data. One can compute various moments of this distribution. For instance, the average time of nucleation conditioned on our data is:
\be
\langle t \rangle = \int_0^\infty {\rm d}t \, t P(t | D ) = \frac{1}{P(D)}\sum\limits_{i \subset I_D}  \int_0^\infty {\rm d}t \, t P(i,t) \,.
\label{mean existence time}
\ee
This is just the mean nucleation time for vacua compatible with our data, weighted by the probability for each.

\subsection{Hypothesis testing and posterior odds}
\label{bayes factor sec}

The Bayesian framework allows one to compare the plausibility of different hypotheses/models to explain the data through the Bayes factor. As mentioned already, we will be specifically
interested in comparing two hypotheses differing only in their priors~$\rho(t)$ for the time of existence: a hypothesis~${\cal H}_{\rm uni}$ with quasi-uniform~$\rho(t)$ ; and a
hypothesis~${\cal H}_{\rm late}$ with a volume-weighted or late-time prior. 

The Bayesian evidence for each hypothesis is given by
\be
P(D | {\cal H}) = \sum_{i \subset I_D} P(i |{\cal H}) \,,
\label{Bayes evidence}
\ee 
where~$P(i |{\cal H})$ is given by~\eqref{predictive distribution} with priors corresponding to~${\cal H}$.   
The relative plausibility of different hypotheses to account for the data is quantified by the {\it Bayes factor}:
\be
\boxed{\frac{P({\cal H}_{\rm late} | D)}{P({\cal H}_{\rm uni} | D)} = \frac{P(D | {\cal H}_{\rm late})}{P(D | {\cal H}_{\rm uni})}\frac{P({\cal H}_{\rm late})}{P({\cal H}_{\rm uni})}}\,,
\label{Bayes factor}
\ee
where~$P({\cal H})$ is the prior for each hypothesis, and~$\frac{P(D | {\cal H}_{\rm late})}{P(D | {\cal H}_{\rm uni})}$ is the {\it posterior odds}.
Assuming equal priors, the Bayes factor coincides with the posterior odds. 

\section{Prior information}

The crux of deriving a measure lies in the specification of prior probability distributions~$p_\alpha$ and~$\rho(t)$.
The choice of priors is a long-standing problem in probability theory, for which there is unfortunately no universally applicable rule.
The following general guiding principles have proven fruitful in other contexts.

Consistency requires that priors reflect {\it all} information at hand. In the case of interest, this includes state-of-the-art understanding of the string
landscape, quantum cosmology {\it etc.} At the same time, our priors should be {\it minimally informative}. They should incorporate all available information, but
should not otherwise be constrained by biases or prejudices. In practice, this is often achieved by applying the principle of indifference or, more generally, the principle of maximal entropy~\cite{jaynes03}.
Lastly, our priors should be {\it objective}, in that two physicists presented with the same information should agree upon a set of justified priors. 

Fortunately, the problem at hand is sufficiently simple and symmetric that the set of justified priors, we will argue, can be narrowed down
to essentially just two possibilities. 
 
\subsection{Prior distribution for the ancestral vacuum}
\label{anc vac prior}

The prior distribution~$p_\alpha$ for the ancestral vacuum pertains to the question of initial conditions. The question of the initial state in quantum cosmology
remains an open problem and has been the subject of active debate for decades. Well-motivated and well-studied proposals include the Hartle-Hawking state~\cite{Hartle:1983ai}
and the tunneling wavefunction~\cite{Vilenkin:1984wp,Vilenkin:1986cy}. At this point, even the qualitative question of whether the initial state should preferentially have
high-entropy/low-energy or low-entropy/high-energy remains unanswered.

For these reasons, it seems prudent to err on the side of maximal ignorance and apply the principle of indifference:
\be
p_\alpha = \frac{1}{N_{\rm dS}}\,.
\label{anc uni}
\ee
Three comments are in order. Firstly, since high-energy dS vacua vastly outnumber low-energy vacua, a uniform prior is statistically equivalent to a prior favoring high-energy/low entropy
initial conditions. Secondly, we will see that our results are highly insensitive to this prior. For all but a very special subset of initial conditions, we will argue that the uniform-time
hypothesis is exponentially preferred over the late-time hypothesis. Lastly, the number of dS vacua may well be infinite in the string landscape~\cite{Silverstein:2001xn,Maloney:2002rr,DeLuca:2021pej}, in which case~\eqref{anc uni} would represent an improper prior. This is of no concern as the resulting probabilities would nevertheless be well-defined.

The master equation also implies that the initial time~$t_0$, which we set to be~$0$, can be traded for a different set of~$p_{\alpha}$, as one can evolve backward or forward in time to get a different initial distribution in the landscape. This ambiguity is not a big issue as we are comparing the two hypotheses,~${\cal H}_{\rm uni}$ and~${\cal H}_{\rm late}$, under the same prior in ancestral vacua~$p_{\alpha}$. 
For the sake of generality we will leave~$p_\alpha$ arbitrary in our analysis below, though we implicitly have in mind the uniform prior~\eqref{anc uni}.

\subsection{dS isometries and prior density for the time of existence} 
\label{time priors} 
 
Applying the principle of indifference (or maximal entropy) to a continuous variable is tricky, simply because a uniform prior for a continuous variable is not reparametrization
invariant. An obvious strategy is to work with a discretized time variable. We will pursue this in Appendix~\ref{discrete time deriv}, and show that the resulting measure agrees exactly with the
continuous-time approach developed here. This offers a non-trivial check on the time-reparametrization invariance of the measure.

Sticking with continuous time, it is instructive to consider the symmetries of the problem and apply the notion of group invariance. As argued by Jaynes~\cite{jaynes03}, one must first identify all transformations that leave our state of knowledge unchanged. Consistency then requires that the prior probabilities be invariant under those transformations. For instance, if one's prior state of knowledge is oblivious to a spatial translation~$x \rightarrow x + c$, then the prior distribution for~$x$ should be uniform. If one's state of knowledge is instead invariant under a scale transformation~$x \rightarrow \lambda x$, then the appropriate choice should be the Jeffreys prior,~$\rho(x) \sim 1/x$. 

In our case we are guided by the symmetries of the master equation and the isometries of dS space. Consistency requires that~$\rho(t)$ be invariant under these transformations. As we will see, these considerations lead to two possible priors, which reflect the well-known dichotomy between local and global approaches to the measure problem. 

\begin{itemize}

\item {\bf Time-translation invariance and uniform-time prior:} As mentioned earlier, a key observation is that the master equation~\eqref{master} is time-translation invariant. More precisely it is invariant under translations of any time variable~$t$ related to proper time via a lapse function~${\cal N}_I$ that is a function of~$H_I$ only. Without any additional information, the uniform prior is the most intuitive and reasonable choice.

Let us warm up using a simple but well known example: the diffusion problem of a particle in one dimension. Given~$P(x|t)$ as the solution to the diffusion equation, which is time-translation invariant, one may ask for the posterior probability $P(t|x)$, which is the distribution of lapsed time~$t$ given that the particle is observed at location~$x$. Without further information, the most intuitive way to calculate such quantity is simply 
\be
P(t|x) = \frac{P(x|t)}{ \frac{1}{T}\int\limits_0^{T} {\rm d}t' P(x|t')}\, .  
\ee
By consulting Bayes' theorem,~$P(t|x)P_x(x)  = P(x|t)P_t(t)$, the intuitive answer corresponds exactly to using a uniform prior with a cutoff in time~$T$, which can be sent to infinity. The prior~$P_t(t)$ in this simple example can be regarded as the distribution of the time of observation. For instance, if instead one were told that most of the observations were concentrated within a particular time window, it would be appropriate in this case to use a non-uniform~$P_t(t)$. 

The Markov process described by the master equation $ \dot{f}_I = \sum_J {\mathbb M}_{IJ}f_J$ is analogous to the simple diffusion problem. Given a distribution of initial condition $p_{\alpha}$, the solution to the master equation, 
\be 
P(I|t) = f_I(t) = \sum_{\alpha} \left({\rm e}^{{\mathbb M}t}\right)_{I\alpha} p_{\alpha}\,,
\label{cond Green}
\ee
is exactly the conditional probability of being in vacuum~$I$ given time~$t$, as explained in the previous Sections. To find the distribution of  lapsed time given that the state is observed in vacuum~$I$,~$P(t|I) =\frac{P(I |t)P_t(t) }{P(I)}$, without additional information, the most natural choice of prior~$\rho(t) =P_t(t)$ with maximal entropy is a uniform prior:
\be
\rho(t) = {\rm constant}\,.  
\label{uni}
\ee
Of course, the uniform distribution on the half real line is not normalizable, and a regularization is needed. A mathematically convenient choice is
\be \label{uni reg}
  \rho(t) = \epsilon {\rm e}^{-\epsilon t}\,,
\ee
with a cutoff~$\sim \epsilon^{-1}$ in time, therefore
\be
  P(t|I) =\frac{P(I|t) \,{\rm e}^{-\epsilon t} }{\int\limits_0^\infty  {\rm d}t' P(I|t')\,  {\rm e}^{-\epsilon t'}}\,.
\ee
At the end of the calculation we will send~$\epsilon\rightarrow 0$ and obtain well-defined, gauge-invariant posterior probabilities.\footnote{One should stress that the nature of~$\epsilon$ is quite different than the late-time cutoff usually introduced to regularize ratios of number counts, such as~\eqref{prob freq}. In our approach~$\epsilon$ is only necessary to make the improper uniform prior well-defined in the intermediate steps. Importantly, none of our results are sensitive to the choice of regulator. For instance, choosing~$\rho(t) = 1/T$ over a large but finite interval~$0 \leq t \leq T$, yields identical results in the limit~$T\rightarrow \infty$.} 

A geometric way to interpret this prior is to focus on one of the metastable dS vacua in our past, for instance our parent vacuum. 
Assuming it is sufficiently long-lived, the parent geometry is approximately invariant under the 10 dS isometries.
Taking a random set of world-lines with initial condition~$p_\alpha$, the chance that a randomly chosen world-line is in vacuum~$I$ at time~$t$ is exactly predicted by the solution to the master equation. 

The action for local operators on the world-line is of the general form
\be
S = \int {\rm d}\tau \,{\cal O}(X^\mu(\tau))\,,
\ee
where the measure~${\rm d}\tau$ is proper time along the curve. After sufficient time, the world-line approaches a geodesic comoving~($\vec{x} = {\rm const.}$) in the dS flat slicing:
\be
{\rm d}s^2 =  - {\rm d}t^2 + {\rm e}^{2Ht} {\rm d}\vec{x}^2 \,.
\label{dS flat}
\ee
Hence~$\tau$ coincides with cosmic proper time~$t$, and the resulting measure~${\rm d}\tau \simeq {\rm d} t$ on the world-line is time-independent. This simple line of reasoning suggests that the prior density~$\rho(t)$ be (proper) time-translation invariant. Note that the prior is also quasi-uniform in e-folding time,~$\rho(N) = \frac{\epsilon}{H} {\rm e}^{- \frac{\epsilon}{H}N}$. In terms of conformal time, it corresponds to the Jeffreys prior,~$\rho(\eta) \sim \eta^{-1}$, consistent with the dS dilation symmetry~$\eta \rightarrow \lambda \eta$,~$\vec{x}\rightarrow\lambda\vec{x}$. 

Heuristically, a time-translation invariant prior reflects complete ignorance about the time of nucleation, akin to a temporal Copernican principle. We simply do not know when eternal inflation started along our particular past world-line. Correspondingly, this is the {\it minimally informative} prior for the time of nucleation.

\item {\bf Volume-weighted prior:} An alternative approach is to consider a finite spatial region in the parent dS geometry. The invariant measure for observables within the region is the usual volume element:
\be
\sqrt{-g}\, {\rm d}t\, {\rm d}^3x = a^3(t) \,{\rm d}t\,{\rm d}^3x\,.
\ee
Averaging over~$\vec{x}$ yields a measure~$\sim a^3(t) \,{\rm d}t$ that grows with volume. This line of reasoning suggests that the
prior density~$\rho(t)$ should similarly grow with volume:
\be
\rho(t) \sim a^3(t)\,.
\label{rho volume}
\ee
This choice can also be motivated heuristically as follows. A time translation in dS corresponds to an exponential growth in volume, and therefore an exponential growth in the number of observers.
If one abides by the self-indication assumption~\cite{bostrom}, whereby prior probabilities are weighted by the number of observers produced, then~$\rho(t)$ should increase exponentially
in time.\footnote{The self-indication assumption can lead to absurd conclusions in other contexts, such as the ``presumptuous philosopher" problem~\cite{bostrom}. Neal instead advocates applying what he
calls Full Non-Indexical Conditioning~\cite{Neal:2006py}. We will argue in Sec.~\ref{model comp} that the late-time hypothesis is disfavored by the data, hence debating the relative merits of these assumptions in eternal inflation is perhaps moot.}

For the prior distribution to be normalizable, a regulator is once again necessary. This can be achieved simply by imposing a cutoff time~$t_{\rm c}$:
\be
\rho(t) = \frac{a^3(t)}{\int\limits_0^{t_{\rm c}} {\rm d}t'\, a^3(t')}  \qquad \text{for}~~0\leq t \leq t_{\rm c} \,,
\label{prior late}
\ee
and~$\rho(t) = 0$ for~$t > t_{\rm c}$. As in the previous case, we will remove the regulator~($t_{\rm c}\rightarrow\infty$) at the end of the calculation and obtain well-defined, gauge-invariant posterior probabilities. 
In fact, we will see that {\it any} growing~$\rho(t)$ that dominates after the relaxation time scale of the landscape gives the same probabilities.

Physically, since~$\rho(t)$ multiplies the occupational probabilities~$f_j(t)$ (see~\eqref{predictive distribution}), this choice amounts to {\it weighing probabilities by physical volume}.
In standard approaches to the measure problem, volume-weighing leads to exponential sensitivity on the choice of time variable~\cite{Linde:1993nz,GarciaBellido:1993wn,Garriga:1997ef}.
In contrast, because volume-weighing is implemented through the prior density~$\rho(t)$ in our context, the resulting probabilities will be manifestly gauge invariant. 

Since the physical volume of all bubble universes grows in proper time asymptotically as~$\sim {\rm e}^{3H_{\rm max}t}$, where~$H_{\rm max}$ is the highest Hubble
rate of any dS vacuum in the landscape~\cite{Aryal:1987vn,Linde:1993xx,Garriga:1997ef},~\eqref{rho volume} is effectively equivalent to, in proper time,
\be
\rho(t) \sim {\rm e}^{3H_{\rm max}t} \,.
\label{prior late proper}
\ee
And because~$H_{\rm max}$ is enormous compared to transition rates governing the evolution of the~$f_j(t)$'s,~\eqref{prior late proper} is well-approximated by a delta function
\be
\rho(t) \simeq 2\, \delta(t-t_{\rm c}) \,;\qquad 0\leq t \leq t_{\rm c}\,.
\label{rho delta}
\ee
Thus, because the prior is so sharply peaked, it is {\it maximally informative}. 
We will make this statement precise below once we derive the prior predictive probabilities in this case. 
The fact that~$\rho(t)$ is sharply peaked at the cutoff time is related to the so-called ``youngness paradox"~\cite{Guth:2007ng} afflicting global measures based
on the proper time cutoff. No such paradox arises in our framework, as far as we can tell.

\end{itemize}

The two justified prior densities derived above are consistent with the assumptions implicit in most approaches to the measure problem. The volume-weighted prior~\eqref{prior late}
is reflected in measures based on the late-time, quasi-stationary distribution~\cite{Linde:1993nz,Linde:1993xx,GarciaBellido:1993wn,Vilenkin:1994ua,Garriga:1997ef,Garriga:2005av}.
For this reason, we refer to this case as the {\it late-time hypothesis}~${\cal H}_{\rm late}$. Meanwhile, the uniform prior~\eqref{uni} is representative of local measures~\cite{Bousso:2006ev,Bousso:2009dm,Bousso:2010zi,Nomura:2011dt}. It is also consistent with the early-time approach to eternal inflation~\cite{Khoury:2019yoo,Khoury:2019ajl,Kartvelishvili:2020thd,Khoury:2021grg,Khoury:2021zao}, which postulates that we exist well-before the exponentially-long mixing time for the landscape. 
We henceforth refer to this prior as {\it uniform-time hypothesis}~${\cal H}_{\rm uni}$.

\section{Prior predictive distributions}
\label{prior predictive sec}

With the above priors at hand, we are now in a position to calculate probabilities relevant to Bayesian inference. In this Section we focus on the prior predictive distribution~$P(I)$,
which gives the probability to occupy vacuum~$I$ irrespective of the time of existence or ancestral vacuum. We will find that the resulting probabilities~$P(I|{\cal H}_{\rm uni})$
and~$P(I|{\cal H}_{\rm late})$ coincide with two measures proposed in the literature, respectively the holographic prior probabilities~\cite{Bousso:2006ev} and the quasi-stationary measure of GSVW~\cite{Garriga:2005av}.
Furthermore, each has a close analogue among centrality indices studied in network science.

\subsection{Uniform-time prior} 
\label{prior predictive with uniform time prior} 
 
We first consider the uniform prior for the time of existence. Substituting~\eqref{uni reg} into~\eqref{predictive distribution}, and using~\eqref{Green}, we obtain
\be
P(I|{\cal H}_{\rm uni})  = \frac{ \sum\limits_{j} \mathbb{T}_{Ij}(\epsilon) \sum\limits_\alpha  \big(\mathds{1} - T(\epsilon)\big)^{-1}_{j\alpha} p_\alpha }{\sum\limits_{J,k}  \mathbb{T}_{Jk}(\epsilon)\sum\limits_\gamma  \big(\mathds{1} - T(\epsilon)\big)^{-1}_{k\gamma} p_\gamma}\,,
\ee
where~$\mathbb{T}_{Ij}(\epsilon) = \frac{\kappa_{Ij}}{\epsilon + \kappa_j}$. Letting~$\epsilon\rightarrow 0$ to remove the regulator gives
\be
\boxed{P(I|{\cal H}_{\rm uni})  = \frac{ \sum\limits_{j} \mathbb{T}_{Ij} \sum\limits_\alpha  \big(\mathds{1} - T\big)^{-1}_{j\alpha} p_\alpha }{\sum\limits_{k,\beta} \big(\mathds{1} - T\big)^{-1}_{k\beta} p_\beta}}\,,
\label{prior distn uni}
\ee
where we have used~$\sum_J \mathbb{T}_{Jk} = 1$ to simplify the denominator. The above probabilities only depend on branching ratios, and therefore are invariant under time reparametrizations.\footnote{This should be obvious, and to prove it we only need~\eqref{kappa gen} to infer that $\frac{\kappa_{Ij}}{\kappa_j} = \frac{\kappa^{\text{proper}}_{Ij}}{\kappa^{\text{proper}}_j}$.} They coincide with Bousso's ``prior probabilities"~\cite{Bousso:2006ev}, derived following a different line of reasoning, and are also closely related to Garriga and Vilenkin's ``comoving" probabilities~\cite{Garriga:2001ri,Garriga:2005av}. 
In our approach, these probabilities are an inevitable consequence of justified objective reasoning with complete ignorance about the time of existence. 

To be precise, Bousso's holographic measure is the product of two factors. First, one considers a single world-line and studies the ensemble of possible future ``histories" of that world-line. The relative probability of different histories, which Bousso calls ``prior probabilities",  is given by their branching ratio probabilities. This is the first factor in Bousso's measure, and it exactly matches~\eqref{prior distn uni}. Bousso then considers the causal diamond for each world-line in the ensemble, and calculates the fraction of observers making different observations within this causal diamond. The fraction of observers is the second factor in his measure. Because we are interested in prior predictive probabilities, without any anthropic conditioning, this second factor is not relevant for our purposes.

The prior predictive distribution~\eqref{prior distn uni} can be easily understood intuitively. Recall from~\eqref{T branch} that~$\left(\mathds{1} - T\right)^{-1}_{ij}$ is the sum of branching probabilities over all paths connecting~$j$ to~$i$. Thus~$P(I|{\cal H}_{\rm uni})$ is naturally interpreted as the sum over all paths connecting ancestral vacua to vacuum~$I$, weighted by the branching probability for each path and averaged over ancestral vacua. It follows that {\it the probabilities~\eqref{prior distn uni} are maximized for vacua that are well-connected~($\mathbb{T}_{Ij} \simeq 1$) to parent dS vacua~$j$ which themselves are easily accessed.} 

The above probability distribution has a close analogue among centrality indices studied in network science. 
Various centrality indices have been proposed in network theory to quantify which nodes in a graph are, in a suitably
defined sense, most important~\cite{centralityreview}. With uniform prior~$p_\alpha = N_{\rm dS}^{-1}$ over ancestral vacua,~\eqref{prior distn uni}
is similar to the {\it Katz centrality} measure~\cite{Katz1953} on a graph:
\be
\vec{C}_{\rm Katz} = \left( \left(\mathds{1} - \alpha A^\intercal \right)^{-1} - \mathds{1} \right)\vec{e}\,,
\ee
where~$\vec{e} = (1,\ldots,1)$ is a vector with unit entries,~$A$ is the graph adjacency matrix, and~$0< \alpha < 1$ is a so-called attenuation parameter.
Intuitively, Katz centrality favors nodes that are well-accessed from other nodes in the network.

\begin{figure}[htb]
\centering
\includegraphics[height=2.5in]{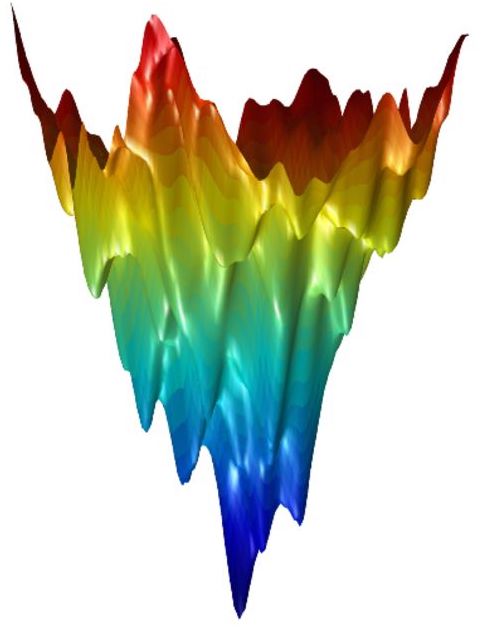}
\caption{The prior predictive probabilities with uniform-time prior favor regions of the landscape where vacua can be accessed through a sequence of downward transitions, from a large basin of ancestral vacua.
Their topography is that of a deep valley, or funnel, akin to the smooth folding funnels of energy landscapes of naturally-occurring proteins. (Reproduced from~\cite{funnelfig}.)}
\label{optimal region}
\end{figure}

\vspace{0.25cm}
\noindent {\bf Funnel topography:} It is also instructive to consider the prior predictive probabilities in the downward approximation (Sec.~\ref{down sec}). Using~\eqref{P down} we have
\be
\sum_\alpha \big(\mathds{1} - T\big)^{-1}_{j\alpha} p_\alpha \simeq p_j + \sum\limits_{\alpha\neq j} {\cal P}_{j\alpha}p_\alpha \equiv p_j + {\cal P}_j \,,
\label{branch ever hit}
\ee
where in the last step we have defined~${\cal P}_j \equiv \sum\limits_{\alpha\neq j} {\cal P}_{j\alpha}p_\alpha$ as the ever-hitting probability to~$j$, averaged over initial conditions.
Hence~\eqref{prior distn uni} becomes
\be
P(I|{\cal H}_{\rm uni})  \simeq \frac{ \sum\limits_{j} \mathbb{T}_{Ij} \left(p_j + {\cal P}_j \right)}{1 + \sum\limits_{k} {\cal P}_k} \,.
\label{prior distn uni down}
\ee
Since this result assumes the downward approximation, the only contributing paths to a given vacuum~$I$ are those given by a sequence of downward transitions. Thus the probabilities~\eqref{prior distn uni down} favor vacua that can be accessed through downward transitions, from a large basin of ancestral vacua. {\it Regions of the landscape with large probability must therefore have the topography of a deep valley, or funnel~\cite{Khoury:2019yoo,Khoury:2019ajl,Kartvelishvili:2020thd,Khoury:2021grg}.} See Fig.~\ref{optimal region}. This is akin to the smooth folding funnels of energy landscapes of proteins~\cite{proteins1}. Naturally-occurring proteins fold efficiently because their free energy landscape is characterized by a smooth funnel near the native state.

\subsection{Late-time/volume-weighted prior} 
\label{prior predict late-time sec}

We next consider the late-time prior~\eqref{prior late} for the time of existence, keeping the prior~$p_\alpha$ over ancestral vacua once again general.
Because the prior is sharply peaked near~$t_{\rm c}$, which is assumed very large, we can approximate the occupational probabilities~$f_j(t)$ by their asymptotic form
\be
f_j(t) \simeq s_j {\rm e}^{-q t}  \,,
\label{fj late}
\ee
where~$s_j$ the dominant eigenvector of~$M_{ij}$ with largest (least negative) eigenvalue~$-q$~\cite{Garriga:2005av}, defined in Sec.~\ref{down sec}.
With~\eqref{prior late} and~\eqref{fj late}, the prior predictive distribution~\eqref{predictive distribution} gives
\be
\boxed{P(I|{\cal H}_{\rm late}) \simeq \frac{ \sum\limits_{j} \kappa_{Ij}  s_j \int\limits_0^{t_{\rm c}} {\rm d} t\, a^3(t) {\rm e}^{-q t}}{\sum\limits_{k} \kappa_{k}s_k \int\limits_0^{t_{\rm c}} {\rm d} t \, a^3(t) {\rm e}^{-q t}} = \frac{\sum\limits_j \kappa_{Ij} s_j }{\sum\limits_{k} \kappa_{k} s_k} }\,.
\label{prior distn late}
\ee
Although we assumed the prior density~\eqref{prior late}, the above makes clear that the result holds for {\it any}~$\rho(t)$ that peaks at sufficiently late times (including the delta function prior~\eqref{rho delta})
such that~\eqref{fj late} is valid. Thus the above probability distribution is manifestly time-reparametrization invariant. It is also independent of the late-time cutoff, such that the limit~$t_{\rm c}\rightarrow \infty$ can be taken. 
Importantly, the result agrees with the GSVW measure~\cite{Garriga:2005av} obtained by counting bubbles along a world-line.\footnote{GSVW implicitly assumes either that this world-line
survives long enough to probe the asymptotically late-time dynamics, or that their measure is the result of averaging over an ensemble of world-lines such that late-time observers dominate the average.} 
To be precise, GSVW obtained, working in e-folding time, 
\be
p_I \sim \sum_j H_j^q  \kappa_{Ij} s_j \,.
\ee
This differs from~\eqref{prior distn late} only by the factor~$H_j^q$, which is indistinguishable from unity since~$q \leq \kappa_\star \lll 1$. 

The prior predictive distribution also admits an intuitive explanation in downward perturbation theory.
Substituting~\eqref{s simple} for the dominant vector,~\eqref{prior distn late} becomes
\be
P(I|{\cal H}_{\rm late}) \simeq  \frac{\sum\limits_j \mathbb{T}_{Ij} \left(\mathds{1} - T\right)^{-1}_{j\star} }{\sum\limits_{k} \left(\mathds{1} - T\right)^{-1}_{k\star} }\,.
\label{prior distn late down}
\ee
The interpretation is clear --- {\it these probabilities are maximized for vacua that are well-connected~($\mathbb{T}_{Ij} \simeq 1$) to parent dS vacua~$j$, which are themselves easily accessed from the dominant vacuum}. Remarkably,~\eqref{prior distn late down} coincides with the uniform-time probabilities~\eqref{prior distn uni} for the special case that~$p_\alpha = \delta_{\alpha\star}$, {\it i.e.}, when
the ancestral vacuum is the dominant vacuum. (This is a manifestation of the local/global duality~\cite{Bousso:2009mw,Bousso:2012cv}.) 
Relatedly, it is often stressed that late-time measures, such as the GSVW measure, reflect the attractor nature of eternal
inflation through their independence on initial conditions. This is certainly the case with~\eqref{prior distn late}. Paradoxically, however, the late-time
distribution~\eqref{prior distn late down} appears to be a special case, corresponding to a particular choice of initial conditions, of the seemingly more general distribution~\eqref{prior distn uni}. 

Because the late-time prior peaks at the cutoff time, it is a maximally informative prior. To see this concretely, consider the probability density~$P(t | I)$ for
the time of existence, conditioned on occupying vacuum~$I$:
\be
P(t | I ) = \frac{P(I,t)}{P(I)}  \,.
\ee
The joint probability is obtained by marginalizing~\eqref{P cond} over~$\alpha$:
\be
P(I,t) \sim \sum\limits_{j} \kappa_{Ij} f_j(t) \rho(t) = \sum\limits_{j} \kappa_{Ij} s_j {\rm e}^{-q t} \rho(t)\,.
\ee
Combining this with~\eqref{prior late proper}, and using the fact that~$H_{\rm max} \gg q$, it is easy to see that
\be
P(t | I ) \simeq \rho(t) \,.
\ee
Thus, even if we had we had complete knowledge of which vacuum we inhabit, we would learn nothing more about the time of
existence than already assumed with this prior.
 
The probability distribution~\eqref{prior distn late} also has a close cousin among centrality indices in network theory.
Namely, it is analogous to {\it eigenvector centrality}, which ranks nodes according to the components of the dominant
eigenvector~$\vec{\pi}$ of the graph's adjacency matrix~$A$.

\section{Model comparison and posterior odds}
\label{model comp}

In the previous Sections we derived, by applying consistent objective reasoning, two hypotheses for probabilities in eternal inflation, which differ only in their priors~$\rho(t)$
for the time of existence: a hypothesis~${\cal H}_{\rm uni}$ with quasi-uniform~$\rho(t)$, which reflects complete ignorance about the time of existence; and a
hypothesis~${\cal H}_{\rm late}$ with volume-weighted~$\rho(t)$, which reflects the belief that we exist at asymptotically late times. In this Section we compare
the plausibility of the two hypotheses to explain our data~$D$ by computing the Bayes factor discussed in Sec.~\ref{bayes factor sec}. 

Substituting the prior predictive probabilities~\eqref{prior distn uni} and~\eqref{prior distn late down}, the Bayesian evidence~\eqref{Bayes evidence} for each hypothesis is
\begin{subequations}
\label{Bayes evidence both}
\bea
\label{Bayes evidence early}
P(D | {\cal H}_{\rm uni})  &=&   \frac{\sum\limits_{i \subset I_D,\, j}  T_{ij} \sum\limits_\alpha  \big(\mathds{1} - T\big)^{-1}_{j\alpha} p_\alpha }{\sum\limits_{k,\beta} \big(\mathds{1} - T\big)^{-1}_{k\beta} p_\beta} \,;\\
\label{Bayes evidence late}
P(D | {\cal H}_{\rm late}) &\simeq &  \frac{\sum\limits_{i \subset I_D,\, j} T_{ij} \left(\mathds{1} - T\right)^{-1}_{j\star} }{\sum\limits_{k} \left(\mathds{1} - T\right)^{-1}_{k\star}} \,,
\eea
\end{subequations}
where the latter is valid in downward perturbation theory. Before attempting to estimate~\eqref{Bayes evidence both}, it is useful to get a sense of the hierarchy of transition rates and branching ratios involved. 


\subsection{Detailed balance and unsuppressed chains}

A generic feature of transition rates in field theory is that they are exponentially staggered. This is because rates depend exponentially on the shape of the potential, such as the height and width of the barrier. Hence branching ratios are typically overwhelmingly dominated by a single decay channel, while other decay channels are comparatively exponentially suppressed.\footnote{There are of course exceptions, for instance in regular lattices of flux vacua~\cite{Bousso:2000xa}, but one expects that single-channel dominance is justified for random landscapes.}

Because of detailed balance, upward transitions are further suppressed by a double exponential factor. To see this, note from~\eqref{detailed balance} that the upward vs downward rate between two vacua with~$V_{\rm low} \ll V_{\rm high}$ is
\be
\frac{\kappa_{\rm up}}{\kappa_{\rm down}} \sim {\rm e}^{-S_{\rm low}}\,.
\label{up down}
\ee
To get a sense of the suppression, for a vacuum with~$V_{\rm low} = (0.1\,M_{\rm Pl})^4$, {\it i.e.}, with energy scale just one order of magnitude
below the Planck scale, the dS entropy is~$S_{\rm low} \simeq {\rm e}^{15}$. A vacuum with our observed~CC,~$V_{\rm low} = 10^{-120}\,M_{\rm Pl}^4$, has~$S_{\rm low} \simeq {\rm e}^{283}$. 
The conclusion is that, if a dS vacuum has one or more available downward decay channels, these will typically have overwhelmingly dominant branching ratios.  

On the other hand, if a vacuum can only decay via upward transitions, then the branching ratio is necessarily dominated by an upward jump.\footnote{In this case the lifetime is of order the dS recurrence time. It has been conjectured that in string theory all dS vacua have a much shorter lifetime~\cite{Freivogel:2008wm}, such that the problem of Boltzmann brains is avoided.}
We will refer to this decay channel as the {\it dominant upward transition}. To gain some intuition on what constitutes a dominant upward transition, 
suppose that a vacuum can either up-tunnel via a high jump to~$V_{\text{high}}$, or via a smaller jump to~$V_{\text{low}}$, once
again with~$V_{\text{low}} \ll V_{\text{high}}$. In this case,~\eqref{detailed balance} implies
\be
\frac{\kappa_{\text{high jump}}}{\kappa_{\text{low jump}}} \sim \frac{\kappa_{\text{large step down}}}{\kappa_{\text{small step down}}} \,{\rm e}^{-S_{\rm low}}\,.
\ee
The prefactor~$\kappa_{\text{large step down}}/\kappa_{\text{small step down}}$ is just the ratio of downward rates to the given vacuum, and its value depends on the details of the potential. It is at best exponentially large and at worse exponentially small. But, in either case, it is generically swamped by the double exponential factor of~${\rm e}^{-S_{\rm low}}$, and thus~$\kappa_{\text{high jump}} \ll \kappa_{\text{low jump}}$. In other words, the dominant upward transition typically corresponds to the smallest increase in potential energy.

These considerations lead us to define an {\it unsuppressed chain} as a sequence of transitions comprised exclusively of downward transitions (whenever these are available) and/or dominant upward transitions (whenever a vacuum can only decay by jumping upwards). In other words, an unsuppressed chain excludes subdominant upward jumps, whose branching ratio is suppressed by a double exponential. And because downward channels dominate whenever they are available, unsuppressed chains consist mostly of downward transitions. We briefly note a few properties of unsuppressed chains:

\begin{itemize}

\item There is at least one unsuppressed chain starting from any vacuum, but any number of them (including zero) can arrive at that vacuum.

\item If an unsuppressed chain exists from~$j$ to~$i$, then~$\left(\mathds{1} - T\right)^{-1}_{ij}$ is at worse exponentially suppressed; if no unsuppressed chain exists, then~$\left(\mathds{1} - T\right)^{-1}_{ij}$ is doubly-exponentially suppressed. 

\item If an unsuppressed chain exists from~$j$ to~$i$, in general no such chain exists from~$i$ to~$j$. 

\end{itemize}
With this intuition and definitions at hand, we can now examine more closely the Bayesian evidence~\eqref{Bayes evidence both} for each hypothesis.

\subsection{Evidence for late-time hypothesis} 

Let us first examine the Bayesian evidence for the late-time hypothesis, given by~\eqref{Bayes evidence late}. Clearly the answer depends on the nature of the dominant
vacuum and its surrounding landscape, which are of course unknown. However, given~$\star$'s status as the most stable vacuum {\it anywhere} in the landscape, it is safe to assume that:~$i)$~it can only decay via an upward transition (because upward jumps are doubly-exponentially suppressed); and $ii)$~it has very small potential energy (because the upward rate is suppressed by~${\rm e}^{-S_\star}$). These assumptions are not new and have been made in other studies of the GSVW measure, {\it e.g.},~\cite{Olum:2007yk,DeSimone:2008if}.

From our earlier discussion, the dominant upward transition from~$\star$ likely proceeds in the direction of smallest increase in potential energy. Following this initial jump,
the unsuppressed chains that emanate from~$\star$ proceed as sequences of downward and dominant upward transitions, until they terminate at terminal vacua. The 
denominator of~\eqref{Bayes evidence late},~$\sum_k  \left(\mathds{1} - T\right)^{-1}_{k\star}$, gives the expected number of dS vacua that are visited in the process. 
All we need for our purposes is that this number is greater than unity, since 
\be
\sum_k  \left(\mathds{1} - T\right)^{-1}_{k\star} = \sum_k \big(\delta_{k\star} + T_{k\star} + \ldots\big) > 1\,.
\label{denom ineq}
\ee
Meanwhile, the numerator,~$\sum\limits_{i \subset I_D}\sum\limits_j T_{ij} \left(\mathds{1} - T\right)^{-1}_{j\star}$ gives the branching probability to reach any vacuum compatible with our data starting from~$\star$. However, because such vacua are (presumably) rare in the landscape, it is highly unlikely that such a vacuum lies along an unsuppressed chain from~$\star$. Instead, as argued in~\cite{Olum:2007yk}, the most probable path to our vacuum likely requires additional (subdominant) upward jumps, such that~$\sum\limits_{i \subset I_D}\sum\limits_j T_{ij} \left(\mathds{1} - T\right)^{-1}_{j\star} \lesssim {\rm e}^{-{\rm e}^{Q}}$, with~${\cal O}(10) \lesssim Q \lesssim  {\cal O}(100)$. Combined with~\eqref{denom ineq}, this implies
\be
P(D | {\cal H}_{\rm late})   \lesssim {\rm e}^{-{\rm e}^{Q}} \,.
\label{PD late final}
\ee
{\it Therefore, in all likelihood,~$P(D | {\cal H}_{\rm late})$ is doubly-exponentially suppressed.} 

\subsection{Evidence for uniform-time hypothesis} 

We next turn our attention to the Bayesian evidence~\eqref{Bayes evidence early} for the uniform-time hypothesis. For concreteness, we first assume the uniform prior~\eqref{anc uni} over ancestral vacua, and then discuss how the results generalize to any prior having non-zero support on high-energy/low entropy initial vacua.

With the uniform prior~$p_\alpha = 1/N_{\rm dS}$, the Bayesian evidence~\eqref{Bayes evidence early} becomes
\be
P(D | {\cal H}_{\rm uni})  =  \frac{\sum\limits_{i \subset I_D} \left(\sum\limits_\alpha \left(\mathds{1} - T\right)^{-1}_{i\alpha} - 1\right)}{\sum\limits_{k,\beta} \big(\mathds{1} - T\big)^{-1}_{k\beta}}\,,
\ee
where we have used the trivial matrix identity~$T\left(\mathds{1} - T\right)^{-1} =  \left(\mathds{1} - T\right)^{-1} - \mathds{1}$. It is convenient to define
\be
N_i \equiv \sum\limits_{\alpha} \big(\mathds{1} - T\big)^{-1}_{i\alpha}
\ee
as the effective number of vacua that can reach~$i$. Note that~$N_i > 1$, since a vacuum can trivially reach itself. In general,~$N_i - 1$ can be doubly-exponentially small (if no unsuppressed chain reaches~$i$), exponentially small (if some unsuppressed chains reach~$i$, but all involve at least one step with exponentially small~$T_{ij}$), or even order unity or larger (if some unsuppressed chains reaching~$i$ solely consist of steps with~$T_{ij} \simeq 1$).

In any case, the Bayesian evidence reduces to
\be
P(D | {\cal H}_{\rm uni})  = \frac{\sum\limits_{i \subset I_D} \left(N_i - 1\right)}{\sum\limits_k N_k}\,.
\label{PD uni N}
\ee
Since~$N_k > 1$, the denominator satisfies~$\sum_k N_k > N_{\rm dS}$. However, since the vast majority of vacua are high-energy vacua, and as such can be reached by few ancestors with non-negligible branching ratio, we expect that~$\sum_k N_k$ does not greatly exceed~$N_{\rm dS}$. That is,
\be
\sum_k N_k \gtrsim N_{\rm dS}\,.
\ee
Meanwhile, the numerator counts the total effective number of vacua that can reach vacua compatible with our data, other than themselves.
Since vacua in~$I_D$ all have a tiny CC of~$\simeq 10^{-120}\,M_{\rm Pl}^4$, it stands to reason that a significant fraction can
be accessed by many vacua via unsuppressed chains. That is,
\be
\sum\limits_{i \subset I_D} \left(N_i - 1\right) \sim N_D\,,
\label{num ND}
\ee
where~$N_D = \sum\limits_{i \subset I_D} N_i$ is the total number of dS vacua compatible with our data. The prefactor in~\eqref{num ND} is at worse exponentially small, but it could also be~$\gg 1$. (This
would be the case is a significant fraction of vacua in~$I_D$ can be accessed by many ancestors through sequences of transitions with~$T_{ij} \simeq 1$.) 

It follows that
\be
P(D | {\cal H}_{\rm uni}) \sim \frac{N_D}{N_{\rm dS}}\,.
\ee
Aside from a prefactor which is at worse exponentially small,~$P(D | {\cal H}_{\rm uni})$ is given by the fraction of all dS vacua compatible with our data. This fraction is of course unknown,
but it is reasonable to expect that it is exponentially small, not doubly-exponentially small. For instance, the worst tuning in the Standard Model is the CC. If the underlying CC distribution
is approximately uniform, then a fraction of~$10^{-120} \simeq {\rm e}^{-276}$ of all dS vacua would have a CC consistent with the observed value.
These considerations lead us to conclude that~$P(D | {\cal H}_{\rm uni})$, while likely exponentially small, is {\it not} doubly-exponentially suppressed like~\eqref{PD late final}.

A similar argument applies to more general prior distributions~$p_\alpha$ over ancestral vacua. To simplify the discussion, we work to leading order in the downward approximation. 
Using~\eqref{prior distn uni down}, the Bayesian evidence becomes
\be
P(D | {\cal H}_{\rm uni})  =  \frac{\sum\limits_{i \subset I_D} {\cal P}_i}{1+ \sum\limits_{k} {\cal P}_k}\,.
\label{PD uni down}
\ee 
In the special case~$p_\alpha = 1/N_{\rm dS}$, we have~${\cal P}_i \simeq \frac{N_i-1}{N_{\rm dS}}$, {\it i.e.},~${\cal P}_i$ measures the effective fraction of other vacua that can reach~$i$.
The argument proceeds along similar lines as the uniform case discussed above. Since most dS vacua in the landscape are high-energy vacua, and as such can be reached through
downward transitions from a limited set of higher-energy vacua, we have~$\sum_k {\cal P}_k\ll 1$. Meanwhile, since vacua compatible with our data all have tiny vacuum energy, we expect that
a significant fraction are accessible via a sequence of downward transitions.\footnote{This is where the prior distribution~$p_\alpha$ over ancestral comes in. If, for some reason, the initial
conditions strongly favor low-energy vacua, then accessing vacua compatible with our data may require subdominant upward jumps, in which case the numerator in~\eqref{PD uni down}
would be doubly-exponentially suppressed.} If so, then we are once again led to conclude that~$P(D | {\cal H}_{\rm uni})$  is likely exponentially small, but {\it not} doubly-exponentially suppressed.

\vspace{0.25cm}
The above analysis leads us to infer that the Bayes' factor~$\frac{P(D | {\cal H}_{\rm late})}{P(D | {\cal H}_{\rm uni})}$ in~\eqref{Bayes factor} is doubly-exponentially small. Assuming comparable
priors for the two hypotheses,~$P({\cal H}_{\rm late}) \sim P({\cal H}_{\rm uni})$, we are led to conclude that {\it posterior odds overwhelmingly favor the uniform-time hypothesis}. 

\subsection{Loopholes}
\label{loopholes}

The above argument is not ironclad. We can think of a few loopholes that would invalidate our conclusions:

\begin{itemize}

\item If the dominant vacuum~$\star$ is, miraculously, compatible with our data, this would imply~$P(D | {\cal H}_{\rm late}) \simeq 1$. This would not only invalidate our conclusions, it would, more importantly, resurrect the old dream of string theory predicting a unique vacuum. Although in this case, the hierarchy between time of existence (which is longer than the mixing time) and the inhabitation time (lifetime of our universe) may potentially be a problem.

\item A more plausible loophole is that there exists a vacuum compatible with our data that lies along an unsuppressed chain starting from~$\star$. This would boost~$P(D | {\cal H}_{\rm late})$ to be only exponentially suppressed, and thus comparable to~$P(D | {\cal H}_{\rm uni})$.  

\item Another plausible loophole is that no vacuum compatible with our data can be reached via a sequence of downward transitions, nor more generally via an unsuppressed chain. This may be because of initial conditions, as mentioned earlier, or because vacua like ours are buried in regions of the landscape that are difficult to access. In either case~$P(D | {\cal H}_{\rm uni})$ would be doubly-exponentially suppressed, and thus comparable to~$P(D | {\cal H}_{\rm late})$. 

\end{itemize}

To determine whether any one of these loopholes is valid would require a more detailed and extensive knowledge of the string landscape. Our conclusion rests on current expectations about the string landscape, to the best of our understanding. We believe that, presented with all the information currently at hand about the landscape and eternal inflation, a bookmaker would set the odds overwhelmingly in favor of the uniform-time hypothesis. Probabilities, after all, are nothing but betting odds~\cite{dutch}.

\section{Inferring the time of existence}
\label{time of existence sec}

The last Bayesian operation of interest is parameter inference, discussed in Sec.~\ref{param infer}. We will be primarily interested in using our data~$D$ to infer the time of existence~$t$. 
For concreteness we focus on the uniform-time prior, since it is overwhelmingly favored by the data, but our analysis can be easily be generalized to any other prior of interest.

Using~\eqref{P cond} with quasi-uniform prior~\eqref{uni reg}, the posterior probability~\eqref{Bayes} for the time of nucleation becomes
\be
P(t | D ) = 
\frac{ \sum\limits_{i \subset I_D,\,j}\kappa_{ij}  \sum\limits_{\alpha} p_\alpha \left({\rm e}^{Mt}\right)_{j\alpha} {\rm e}^{-\epsilon t}}{ \sum\limits_{k \subset I_D,\,\ell}T_{k\ell}  \sum\limits_{\beta}  \big(\mathds{1} - T\big)^{-1}_{\ell\beta} p_\beta}\,,
\label{Bayes uni}
\ee
where in simplifying the denominator we have used~\eqref{Green} and sent~$\epsilon\rightarrow 0$. We can compute various moments of this distribution. For instance, the average time of nucleation~\eqref{mean existence time} is
\bea
\nonumber
\langle t \rangle = \int_0^\infty {\rm d}t \, t P(t | D) &=&  \lim_{\epsilon\rightarrow 0} \frac{ \sum\limits_{i \subset I_D,\,j}\kappa_{ij}  \sum\limits_{\alpha} p_\alpha \int\limits_0^\infty {\rm d}t\,t  \left({\rm e}^{Mt}\right)_{j\alpha} {\rm e}^{-\epsilon t}}  {\sum\limits_{k \subset I_D,\,\ell}T_{k\ell}  \sum\limits_{\beta}  \big(\mathds{1} - T\big)^{-1}_{\ell\beta} p_\beta} \\
&=&  -\frac{ \sum\limits_{i \subset I_D,\,j}\kappa_{ij}  \sum\limits_{\alpha} p_\alpha \times \left.\frac{{\rm d}}{{\rm d}\epsilon}(\epsilon - M)^{-1}_{j\alpha}\right\vert_{\epsilon = 0}} {\sum\limits_{k \subset I_D,\,\ell}T_{k\ell}  \sum\limits_{\beta}  \big(\mathds{1} - T\big)^{-1}_{\ell\beta} p_\beta}\,.
\label{mean t 0}
\eea
Using~\eqref{Green} it is straightforward to obtain
\be   \label{eqn:timeexist}
\langle t \rangle = \frac{ \sum\limits_{i \subset I_D,\,j} T_{ij} \sum\limits_m \big(\mathds{1} - T\big)^{-1}_{jm} \kappa_m^{-1} \sum\limits_\alpha\big(\mathds{1} - T\big)^{-1}_{m\alpha} p_\alpha}{\sum\limits_{k \subset I_D,\,\ell}T_{k\ell}  \sum\limits_{\beta}  \big(\mathds{1} - T\big)^{-1}_{\ell\beta} p_\beta}\,.
\ee

In the downward approximation, together with the assumption that transition rates from most low lying dS vacua to AdS terminals are always larger than up-tunneling rates, this expression simplifies to
\be
 \langle  t \rangle \simeq \frac{\sum\limits_{i  \subset I_D }\langle t_i\rangle  }{\sum\limits_{j \subset I_D } {\cal P}_j} \,,
\label{mean t}
\ee
where we have defined an average unconditional MFPT to~$i$ as 
\be
\langle t_{i} \rangle = \sum_{\alpha} \langle t_{i \alpha}\rangle p_{\alpha}  = \sum_{\ell \neq i, \alpha}\kappa_{i \ell } \left( M^{(i)}\right)^{-2}_{\ell j} p_{\alpha}\,.
\ee
The technical details of the derivation is given in Appendix~\ref{app:time}. This result is intuitively clear. The average time to reach vacua~$i$ compatible with our data is the average of the characteristic time~$\langle t_{j}\rangle$ to reach a parent vacuum.

The arguments of Sec.~\ref{model comp}, which lead us to conclude that the uniform-time hypothesis is overwhelmingly favored, relied on the assumption 
that there is at least one vacuum compatible with our data which can be accessed through a sequence of downward transitions. Equation~\eqref{mean t}
also hinges on that assumption. Since downward transition rates are exponentially faster than upward rates (see~\eqref{up down}), the downward MFPT is
correspondingly exponentially shorter than the dS recurrence time of low-energy vacuum, and thus certainly exponentially shorter than the mixing time for the landscape.

Therefore, if vacua compatible with our data can be accessed via downward transitions, we most likely exist at early times in the unfolding of the multiverse, well-before the mixing time for the landscape.
This confirms the assumptions underlying the early-time approach to eternal inflation~\cite{Khoury:2019yoo,Khoury:2019ajl,Kartvelishvili:2020thd,Khoury:2021grg}. This also circumvents the issue of Boltzmann brains~\cite{Albrecht:2002uz,Dyson:2002pf,Albrecht:2004ke,Page:2005ur,Page:2006dt}, which are produced on exponentially longer time scales. On the other hand, the time of existence for the late time hypothesis  is much longer than mixing time and recurrence time of certain vacua. The rate of production of freak observers put stringent constraint on the landscape.

\section{Conclusion}

Understanding our place within this multiverse is ultimately necessary to make {\it any} predictions about physical observables in our universe.
Attempts to define probabilities (or measure) usually rely on limiting frequency distributions. This is perhaps natural, since the infinite ensemble necessary
to define frequencies is actually realized in the multiverse. Unfortunately, this approach has failed to yield an unambiguous answer. 

In this paper we instead applied Bayesian reasoning to define probabilities. The advantage of this approach is first and foremost a practical one. All attempts to define
a semi-classical measure rely on certain assumptions. The Bayesian framework naturally compels one to make all assumptions explicit through prior information.
Our approach has been strongly influenced by Jaynes' view of probability theory as an extension of classical logic~\cite{jaynes03}. Probabilities, in this viewpoint, amount to reasonable
expectations~\cite{Cox} drawn from limited information. Our treatment is also inspired by Caves' elegant resolution~\cite{Caves:2000tx} to the Doomsday argument~\cite{doom1,doom2,doom3}.
Our approach does not rely on ad hoc geometric constructions, nor are we counting anything.

The natural starting point to define probabilities is the master equation governing vacuum dynamics, obtained after suitable coarse-graining. Remarkably,
this equation describes a linear Markov process, free of the conceptual pitfalls of eternal inflation. This is not a mathematical artifice --- physically, the master
equation describes the random walk on the network of vacua that ``we" have performed since the onset of eternal inflation. The occupational probabilities~$f_I(t)$'s 
are normalized and time-reparametrization invariant, and thus offer well-defined probabilities to occupy different vacua at time~$t$.

Our probabilities require two pieces of prior information: a prior probability density~$\rho(t)$ for the time of nucleation; and a prior probability~$p_\alpha$ for the ancestral vacuum.
Both pertain to initial conditions. We know that eternal inflation started a finite time in our past, but we do not know when. And it started in some particular vacuum, but we do not
know which one. Different approaches to the measure problem amount to different choices for these priors.

Consistency requires that our priors reflect all information at hand, but should otherwise be minimally informative.
For ancestral vacua, we advocated the uniform prior~$p_\alpha = 1/N_{\rm dS}$ as a conservative choice, though our
conclusions are fairly insensitive to this choice. What matters is that the initial conditions have support over
high-energy/low-entropy vacua. For the time of nucleation, we argued that a quasi-uniform prior is a natural choice, consistent with the time-translational
invariance of the master equation. It represents the minimally-informative prior. The resulting predictive probability distribution
matches the prior probabilities of~\cite{Bousso:2006ev} and is closely related to the ``comoving" probabilities discussed in~\cite{Garriga:2001ri,Garriga:2005av}.

We also considered a volume-weighted~$\rho(t)$, which amounts to weighing probabilities for different vacua by their physical volume.
This prior peaks at late times, and as such is maximally informative. Interestingly, because volume weighing is implemented as a prior, the resulting
probabilities do not suffer from the usual sensitivity to the choice of time variable and the associated paradoxes. Instead, the predictive
distribution is time-reparametrization invariant, and agrees with the GSVW measure~\cite{Garriga:2005av}. 

The Bayesian framework allowed us to compare the plausibility of the uniform-time and volume-weighted hypotheses to explain our data
by computing the Bayesian evidence for each. We argued, under general and plausible assumptions, that posterior odds overwhelmingly
favor the uniform-time hypothesis. The argument relies on assumptions about the dominant vacuum that have been made in
previous studies of the landscape, {\it e.g.},~\cite{Olum:2007yk}. There are some caveats, of course, and we tried to enunciate
them carefully in Sec.~\ref{loopholes}. 

We believe that the uniform-time measure is the correct objective approach to probabilistic reasoning in the multiverse.
The assumed priors are the least informative and consistently reflect our current state of knowledge about how/when eternal
inflation started. The prior predictive distribution~\eqref{prior distn uni} is very intuitive, to the extent that one could have written
down the answer without doing any work. It favors vacua that are easily accessed under the random walk on the landscape.

Despite making the least informative prior assumptions, the probability distribution~\eqref{prior distn uni} is surprisingly predictive. It favors vacua lying within deep funnels~\cite{Khoury:2019yoo,Khoury:2019ajl,Khoury:2021grg}, wherein they can be accessed through a sequence of downward transitions from a large basin of parent vacua. This is akin to the folding funnels of proteins~\cite{proteins1}. As argued in Sec.~\ref{time of existence sec}, it also predicts that we exist at times much earlier than the mixing time for the landscape, confirming the intuition behind the early-time approach to eternal inflation~\cite{Khoury:2021grg}. This implies, incidentally, that we are ``normal" observers as opposed to Boltzmann brains~\cite{Albrecht:2002uz,Dyson:2002pf,Albrecht:2004ke,Page:2005ur,Page:2006dt}, which are produced on exponentially longer time scales. 

The analogy with natural selection and protein landscapes is quite apt. The prior predictive probabilities~\eqref{prior distn uni} define a fitness function on the string landscape, analogous to the fitness landscape over the space of protein sequences~\cite{Wright1932}. Sequence space is high dimensional, much like the string landscape. And it is believed that, through evolution, {\it all} of sequence space for proteins and genomes has been explored by biology on Earth~\cite{sequencespace}, just like eternal inflation is efficient at populating the entire landscape. Naturally-occurring proteins occupy a small region of the fitness landscape, characterized by a large basin of ``neutral mutations"~\cite{superfunnelsrev}. These large basins with high fitness, known as ``superfunnels"~\cite{superfunnels1,superfunnels2,superfunnels3}, are strikingly similar to the large funnels favored by our measure.

The probabilistic framework developed here opens up many avenues of inquiry. We mention in closing two particularly interesting directions:

\begin{itemize}  
 
\item By modeling landscape regions as random networks, we will show in a forthcoming paper~\cite{future} that the uniform-time probabilities~\eqref{prior distn uni} favor regions that are close to the 
directed percolation phase transition~\cite{Odor:2002hk}. Thus landscape dynamics belong to the universality class of directed percolation --- the paradigmatic non-equilibrium critical phenomenon. As usual, the
predictive power of criticality lies in scale invariant observables characterized by critical exponents. We will argue that the probability distribution for the CC is a power-law,
that favors a naturally small and positive vacuum energy. Tantalizingly, this hints at a deep connection between non-equilibrium critical phenomena on the landscape and the near-criticality of our universe.

\item The connection with protein folding funnels deserves further exploration. The problem of search optimization on complex, high-dimensional energy landscapes has already been solved by naturally-occurring proteins. A fascinating result in protein folding is that conformation networks share many properties of real-world networks~\cite{Rao_Caflisch_2004}: their degree distribution is scale-free, they enjoy the small-world property, and they are hierarchical. It will be interesting to study the implications of similar properties holding in regions of the string landscape.

\end{itemize}

\vspace{.4cm}
\noindent
{\bf Acknowledgements:} We thank Raphael Bousso, Dick Bond, Cliff Burgess, Paolo Creminelli, Giorgos Gounaris, Alan Guth, James Halverson, Oliver Janssen, Eleni Katifori, Mehrdad Mirbabayi, Miguel Montero, Yasunori Nomura, Federico Piazza, Eva Silverstein, Henry Tye, Cumrun Vafa, Alex Vilenkin and Elizabeth Wildenhain for helpful discussions. We thank Bjoern Friedrich, Arthur Hebecker, Manfred Salmhofer, Jonah Strauss and Johannes Walcher for enlightening correspondence on their Wheeler-de Witt approach~\cite{Friedrich:2022tqk}. This work is supported by the US Department of Energy (HEP) Award DE-SC0013528, NASA ATP grant 80NSSC18K0694, and by the Simons Foundation Origins of the Universe Initiative.

\begin{appendices}

\section{Discrete-time derivation}
\label{discrete time deriv}

In this Appendix we present an alternative derivation of the prior predictive distribution~\eqref{prior distn uni} using a discrete time variable. The latter is defined such
that each dS vacuum undergoes a transition (to another dS or to a terminal) at every time step~$n = 0,1,2,\ldots$~\cite{Garriga:2012bc}. 

The probability~$X_I(n)$ to occupy vacuum~$I$ at time~$n$ satisfies the master equation
\be
X_I(n+1) = \sum_J \mathbb{T}_{IJ}X_J(n)\,,
\label{master discrete}
\ee
where the full branching matrix~$\mathbb{T}_{IJ}$ was defined in~\eqref{branch mtx}. Its sum
rule~\eqref{branch mtx sum} ensures that probability is conserved
\be
\sum_I X_I(n) = 1\,.
\ee
The solution to~\eqref{master discrete} is given by
\be
X_I(n) = \sum_\alpha \big(\mathbb{T}^n\big)_{I\alpha} p_\alpha \,,
\label{X soln}
\ee
where~$p_\alpha \equiv X_\alpha(0)$ denotes as before the initial probability over ancestral vacua. For dS vacua, in particular,~\eqref{X soln} becomes
\be
X_j(n) = \sum_\alpha \big(T^n\big)_{I\alpha} p_\alpha \,.
\label{X soln dS}
\ee

The joint probability~$P(I,n)$ to inhabit vacuum~$I$ after a nucleation event at the~$n^{\rm th}$ time step is
\be
P(I,n) = {\cal N} \sum_j \mathbb{T}_{Ij} X_j(n-1) P_{\rm prior}(n)  \,,
\ee
where~${\cal N}$ is a normalization constant. This is the discrete-time analogue of~\eqref{P cond}, after marginalizing over~$\alpha$,
with the prior~$P_{\rm prior}(n)$ on the discrete time of existence playing the role of~$\rho(t)$. The prior predictive distribution is then given by
\be
P(I) = \sum_{n=0}^\infty  P(I,n) P_{\rm prior}(n) = \frac{\sum\limits_j \mathbb{T}_{Ij} \sum\limits_{n=1}^\infty X_j(n-1) P_{\rm prior}(n)}{ \sum\limits_k \sum\limits_{m=1}^\infty X_k(m-1) P_{\rm prior}(m)} \,,
\label{prior pred discrete}
\ee
where we have fixed the normalization constant to ensure that~$\sum\limits_I P(I) = 1$. Note that the sum over~$n$ starts at~$n = 1$ since, by assumption, there has been at least one nucleation event along our past world-line, {\it i.e.}, the one giving rise to our bubble. 

An improper uniform prior in this case is straightforward and given by~$P_{\rm prior}(n) = {\rm constant}$ for all~$n \geq 1$. This reflects complete prior ignorance about how many nucleation events took place along our past world-line. As in the continuous case, the uniform-time prior must be regularized. A simple prescription is to introduce a large cutoff~$M$:
\be
P_{\rm prior}(n) = \left\{\begin{array}{cl}
 \frac{1}{M}   & ~~n = 1,\ldots, M  \\
0  & ~~\text{otherwise}  \,.
\end{array}\right.
\ee
Substituting this prior, together with~\eqref{X soln dS}, the prior predictive distribution~\eqref{prior pred discrete} becomes 
\be
P(I|{\cal H}_{\rm uni})  = \frac{\sum\limits_{j,\alpha} \mathbb{T}_{Ij} \sum\limits_{n = 1}^{M}  (T^{n-1})_{j\alpha} p_\alpha}{ \sum\limits_{k,\beta} \sum\limits_{m = 1}^{M}  (T^{m-1})_{k\beta} p_\beta} = \frac{\sum\limits_{j,\alpha} \mathbb{T}_{Ij} \sum\limits_{n = 0}^{M-1}  (T^{n})_{j\alpha} p_\alpha}{ \sum\limits_{k,\beta} \sum\limits_{m = 0}^{M-1}  (T^{m})_{k\beta} p_\beta}\,.
\ee
At this point the cutoff can be removed by sending~$M\rightarrow \infty$, with the result
\be
P(I|{\cal H}_{\rm uni})  = \frac{ \sum\limits_{j} \mathbb{T}_{Ij} \sum\limits_\alpha  \big(\mathds{1} - T\big)^{-1}_{j\alpha} p_\alpha }{\sum\limits_{k,\beta} \big(\mathds{1} - T\big)^{-1}_{k\beta} p_\beta}\,.
\ee
This agrees precisely with the continuous-time answer~\eqref{prior distn uni}. It reaffirms that our probabilities are time-reparametrization invariant. 

\section{Exact relation between time of existence and first passage time}
\label{app:time}
We provide an exact relation between the average time of existence~\eqref{eqn:timeexist} and mean first-passage time in this section. The time of existence~\eqref{eqn:timeexist} derived in Sec.~\ref{time of existence sec} can be written in terms of the transition matrix~$M$ as 
\be 
  \langle t \rangle = \frac{1}{ {\cal N} } \sum_{i \subset I_D, j , \alpha  } \kappa_{i j} \Big( M^{-2} \Big)_{j\alpha} p_{\alpha} \,,
\ee
where~${\cal N} =- \sum_{i \subset I_D, j , \alpha} \kappa_{i j} \Big( M^{-1} \Big)_{j\alpha} p_{\alpha}$ is the normalization factor of the conditional probability $P(t|D)$. Recall that the unconditional MFPT~$\langle t_{ij} \rangle$ and the conditional MFPT~$\langle t_{ij} \rangle_c$ from~$j$ to~$i$ are given by  \cite{Khoury:2021grg}
\be
\langle t_{ij} \rangle  = \sum_{l\neq i}\kappa_{i l} \left( M^{(i)}\right)^{-2}_{lj}\,; \quad  \langle t_{ij} \rangle_c  = \frac{\sum_{l\neq i}\kappa_{i l} \left( M^{(i)}\right)^{-2}_{lj} }{ {\cal P}_{ij} }\,,
\ee
where~$M^{(i)}$ is the transition matrix with the~$i$-th column and row deleted, and~${\cal P}_{ij}=-\sum_{k\neq l}\kappa_{i l} \left( M^{(i)}\right)^{-1}_{lj}$ is the ever-hitting probability. 

To find a relation between~$ \langle t \rangle $ and ~$\langle t \rangle_{ij}$, we can use the relation between~$M^{-1}$ and~$\left(M^{(i)}\right)^{-1}$,
\begin{align}
&M^{-1} = \begin{bmatrix} 
 -\kappa_i    &     \cev{\kappa}_i \\
  \vec{\kappa}_i  & M^{(i)}
\end{bmatrix}^{-1} = \begin{bmatrix} 
- Q_i    &     \cev{\kappa}_i W_i Q_i \\
 Q_iW_i \vec{\kappa}_i  &  (\mathds{1} - Q_i W_i \vec{\kappa}_i\cev{\kappa }_i ) W_i
\end{bmatrix}\,;\nonumber \\  
& Q_i = (\kappa_i + \cev{\kappa}_i W \vec{\kappa}_i)^{-1} = (\kappa_i - \cev{\cal P}_i\vec{\kappa}_i)^{-1} \,; 
\quad W_i =  \left(M^{(i)}\right)^{-1}\,,
\end{align}
where~$[\vec{\kappa}_i]_j = \kappa_{ji}$,~$[\cev{\kappa}_i]_j = \kappa_{ij}$ and~$[\cev{\cal P}_i]_j = {\cal P}_{ij}$ for~$j\neq i$. Note that~$Q_i$ has an interesting meaning when expanded out in the following way,
\be
 Q_i = \kappa_i^{-1}  +  \kappa_i^{-2} \cev{\cal P}\vec{\kappa}_i +  \kappa_i^{-3} (\cev{\cal P}\vec{\kappa}_i )^2+ \ldots\,,
\ee
the $n$-th term $\kappa_i^{-n-1} (\cev{\cal P}\vec{\kappa}_i )^n$ can be interpreted as the time needed for the trip~$i \rightarrow (\mbox{any }j  \rightarrow i \rightarrow )^n\rightarrow a$. 
To compute $\langle t \rangle$ we need 
\be
 M^{-2} = \begin{bmatrix}
  Q^2(1+ \cev{\kappa} W^2 \vec{\kappa} ) &  - \cev{\kappa} W Q^2 + \cev{\kappa}W^2 Q - \cev{\kappa}W^2 \vec{\kappa} \cev{\kappa} W Q^2 \\
  W\vec{\kappa} Q^2 + W^2\vec{\kappa} Q  - W\vec{\kappa}\cev{\kappa}W^2 \vec{\kappa}Q^2 &  Q^2 W\vec{\kappa} \cev{\kappa} W + W^2 - W^2\vec{\kappa} \cev{\kappa} WQ -W\vec{\kappa} \cev{\kappa} W^2  Q + W\vec{\kappa} \cev{\kappa} W^2\vec{\kappa} \cev{\kappa} W  Q^2
 \end{bmatrix}\,.
\ee
Note that we have omitted the subscript~$i$ to avoid clustering. Therefore the time scale~$t_i = \sum_{j,\alpha} \kappa_{ij} \left(M^{-2} \right)_{j\alpha} p_{\alpha}$ can be written as 
\begin{align}
t_i &= \sum_{j,\alpha} \kappa_{ij} \left(M^{-2}  \right)_{j \alpha} p_{\alpha} \nonumber \\
 &= \left( \cev{t}_i+ Q_i^2 \cev{\cal P}_i \vec{\kappa}_i \cev{\cal P}_i  + Q_i\cev{t}_i \vec{\kappa}_i \cev{\cal P}_i  +  Q_i\cev{\cal P}_i \vec{\kappa}_i \cev{t}_i  + Q_i^2 \cev{\cal P}_i \vec{\kappa}_i \cev{t}_i \vec{\kappa }_i \cev{\cal P}_i \right)\cdot p\,,
\end{align}
where~$[\cev{t}_i]_j = t_{ij} $ is a row vector of MFPTs from~$j$ to~$i$. Finally the time of existence~$\langle  t \rangle$ is just the sum over all~$i$ that are compatible with data,
\be
 \langle  t \rangle = \frac{1}{\cal N}\sum_{i\subset I_D} t_i\,.
\ee
Also note that the normalization factor ${\cal N}$ can be written in terms of ever-hitting probabilities,
\begin{align}
{\cal N} &= -\sum_{i \subset I_D, \,j , \,\alpha} \kappa_{i j} \Big( M^{-1} \Big)_{j\alpha} p_{\alpha}  \nonumber\\ 
   &= \sum_{i \subset I_D } \left(  \cev{\cal P}_i+ Q_i \cev{\cal P}_i \vec{\kappa}_i \cev{\cal P}_i  \right)\cdot p \,.
\end{align}
In the downward approximation, either 
\begin{align}
&\left[\vec{\kappa}_i\right]_j = \kappa_{ji} =0\,; \quad   \left[\cev{\cal P}_i \right]_j \ge 0~~\mbox{ ($\neq 0$ when there is a directed path from $j$ to $i$)}  \nonumber \\
\text{or}~~~& \left[\vec{\kappa}_i\right]_j = \kappa_{ji} \geq 0\,; \quad  \left[\cev{\cal P}_i \right]_j  = 0 \quad \mbox{(when there are only upward paths to go from $j$ to $i$)} 
\end{align}
can happen for any node $j$ above or below, therefore the scalar product $\cev{\cal P}_i \vec{\kappa}_i$ goes to zero in the downward approximation. With some effort in analyzing $\cev{t}_i \vec{\kappa}_i$, one finds that it also vanishes in the downward approximation. (This is somewhat counter intuitive as it is an unconditional quantity.) The essential part is that there exist terminals, and the inverse of the matrix is just rational function of its elements. The denominator of $\left(M^{(i)}\right)^{-2}$ is simply $\det(M^{(i)})^2 = \prod_{\neq i} (-\kappa_j)^2 $ in the downward approximation. Given that the transition rate to AdS terminals is faster than the up-tunneling rate, the denominator would not go to zero in the downward approximation. It is also obvious that the numerator should be of $\cal O$(up-tunneling rate) for up-tunneling paths. Therefore in the downward approximation, 
\begin{align}
 \langle  t \rangle = \frac{\sum\limits_{i  \subset I_D }\langle t_i\rangle  }{\sum\limits_{j \subset I_D } {\cal P}_j} \,,
\end{align}
where $\langle t_i\rangle  = \cev{t}_i\cdot p $ and   ${\cal P}_i  = \cev{\cal P}_i\cdot p $.

\end{appendices}

\bibliographystyle{utphys}
\bibliography{bayesian_measure_eternal_inflation_v9.bib}
\end{document}